\documentstyle[12pt,aps,epsfig]{revtex}
\title{\Large\bf QCD string in light-light and heavy-light mesons}
\author{\large Yu.S.Kalashnikova\thanks{yulia@heron.itep.ru  $^{**}$nefediev@cfif.ist.utl.pt
$^{***}$simonov@heron.itep.ru}$^a$, A.V.Nefediev$^{**\;a,b}$,
Yu.A.Simonov$^{***\;a}$}
\address{$^a$Institute of Theoretical and Experimental Physics, 117218,\\ B.Cheremushkinskaya 25, Moscow, Russia}
\address{$^b$Centro de F\'\i sica das Interac\c c\~oes Fundamentais (CFIF),
Departamento de F\'\i sica, \\Instituto Superior T\'ecnico, Av. Rovisco Pais, P-1049-001 Lisboa, Portugal}

\newcommand{\be}{\begin{equation}}
\newcommand{\ee}{\end{equation}}
\newcommand{\ds}{\displaystyle}
\newcommand{\low}[1]{\raisebox{-1mm}{$#1$}}

\newcommand{\lmn}{\mathop{\sim}\limits_{n\gg 1}}
\begin{document}
\maketitle
\bigskip

\centerline{\large\bf Abstract}
\medskip

\begin{abstract}
The spectra of light--light and heavy--light mesons are calculated within the framework of
the QCD string model, which is derived from QCD in the Wilson loop approach. Special attention is payed to the
proper string dynamics that allows us to reproduce the straight-line Regge trajectories
with the inverse slope being $2\pi\sigma$ for light--light and twice as small for
heavy--light mesons.
We use the model of the rotating QCD string with quarks at the ends to
calculate the masses of several light-light mesons lying on the lowest Regge
trajectories and compare them with the experimental data as well as with the
predictions of other models.
The masses of several low-lying orbitally and radially excited heavy--light states in the
$D$, $D_s$, $B$, and $B_s$ meson spectra are calculated in the einbein 
(auxiliary) field approach,
which has proven to be rather accurate in various calculations for relativistic systems.
The results for the spectra are compared with the experimental and recent lattice
data. It is demonstrated that an account of the proper string dynamics encoded in
the so-called string correction to the interquark interaction leads to an extra negative
contribution to the masses of orbitally excited states that resolves the problem of
the identification of the $D(2637)$ state recently claimed by the DELPHI Collaboration.
For the heavy-light system we extract the constants $\bar\Lambda$, $\lambda_1$, and $\lambda_2$ 
used in Heavy Quark Effective Theory (HQET) and find good agreement with the results 
of other approaches.
\end{abstract}
\bigskip
{\parindent=0cm PACS: 12.38.Aw, 12.39.Hg, 12.39.Ki}
\bigskip

\section{Introduction}\label{sec1}

The description of the mass spectrum of hadrons is one of the fundamental problems of strong
interactions. It has been attacked in a sequence of approaches motivated by QCD, but still
attracts considerable attention. One of the most intriguing phenomena --- namely, the
formation of an extended object, the QCD string, between the colour constituents inside
hadrons, --- plays a crucial role in understanding their properties. In the present paper this
role is exemplified by spectra of the mass of light-light and heavy-light mesons. In the
former case we study the role played by the QCD string in formation of the
straight-line Regge trajectories and discuss the form of the interquark interaction inside
light hadrons. For heavy-light mesons we find the masses of several low-lying states
in the $D$, $D_s$, $B$, and $B_s$ meson spectra including orbitally and radially excited ones.

We calculate and discuss the spin-spin and spin-orbit splittings and compare them to 
experimental and recent lattice data. Special attention is payed to the role of the
proper string dynamics in establishing the correct slope of the Regge trajectories for
both light-light and heavy-light states, as opposed to those following from
relativistic equations with local potentials.

We remind the reader then that an extra piece of the effective interquark potential, the string
correction, which is entirely due to the string-type interaction in QCD \cite{MP,DKS},
gives a negative
contribution to the masses of orbitally excited states. The latter observation allows us to
resolve the \lq\lq mystery" of an extremely narrow $D(2637)$ state (and a similar one in the
$B$-mesonic spectrum) \cite{KNn} recently claimed by the DELPHI Collaboration \cite{DELPHI,DELPHI2}. We present a reasonable
fit for the several lowest states in $D$- and $B$-mesonic spectra using the
standard values for the string tension, the strong coupling constant, and the current quark
masses. We also find the correspondence between our model and Heavy Quark Effective
Theory, extracting the constants used in the latter approach in the expansion of a
heavy-light meson mass in the inverse powers of the heavy quark mass. We find
analytical formulae for these constants and compare their numerical estimates with
the predictions of other models.

The two main approaches used in the numerical calculations are the quasiclassical method
of solving the eigenenergies problem and the variational one based on the einbein field
formalism. The accuracy of both methods is tested using exactly solvable equations and found
to be about 7\% at worst even for the lowest states. Possible improvements of the method are outlined and discussed.

The paper is organized as follows. In Section \ref{sec2} we give a brief insight into various
aspects of the einbein field formalism. In Section \ref{sec3} the exact spectra of relativistic
equations are compared to the results of approximate calculations using the
quasiclassical and variational einbein field methods, as well as the combined one.
In Section \ref{sec4} we discuss the problem of the Regge trajectory slopes as they appear from the
relativistic equations with local potentials and from the string-like picture of
confinement.
Derivation of the Hamiltonian for the spinless quark-antiquark system as well as of the
spin-dependent corrections to it is the subject of Section \ref{sec5}. The spectra of light-light and
heavy-light mesonic states are calculated and discussed in Sections \ref{sec6} and
\ref{sec7}, respectively. Section \ref{sec8} contains our conclusions and outlook.

\section{Einbein field formalism}\label{sec2}

In this section we give a short introduction to the method of einbein fields and its
possible applications to relativistic systems. The interested reader can find a more
detailed information in \cite{we-ein} and references therein.

\subsection{Reparametrization invariance and constrained systems}\label{subsec21}

Historically the einbein field formalism was introduced in \cite{einbein} to treat the kinematics of the
relativistic spinless particles. Later it was generalized to the case of spinning
particles \cite{spin} and strings \cite{generalized}. So the action of a free relativistic particle can be rewritten
as\footnote{In the path integral formalism this transformation is based on the following
relation
$$
\int D\mu(\tau)\exp{\left(-\int d\tau\left(\frac{a\mu}{2}+\frac{b}{2\mu}\right)\right)}
\sim\exp{\left(-\int d\tau\sqrt{ab}\right)}.
$$}
\be
S=\int_{\tau_i}^{\tau_f}L(\tau),\quad L=-m\sqrt{\dot{x}^2}\to -\frac{m^2}{2\mu}-
\frac{\mu\dot{x}^2}{2},
\label{L1}
\ee
where the dot denotes derivative with respect to the proper time $\tau$, $\mu$ being the
einbein field\footnote{Usually $e=\frac{1}{\mu}$ is referred to as the einbein \cite{einbein}.}. The original
form of the action can be easily restored after solving the Euler--Lagrange equation of
motion for the einbein $\mu$ which amounts to taking the extremum in the latter. Note that the
invariance of the initial action with respect to the change of the proper time,
\be
\tau\to f(\tau),\quad \frac{df}{d\tau}>0,\quad f(\tau_i)=\tau_i,\quad f(\tau_f)=\tau_f,
\label{gl}
\ee
is preserved if an appropriate rescaling is prescribed to $\mu$:
\be
\mu\to\mu/\dot{f}.
\ee

The latter invariance means that one deals with a constrained system. For the free
particle the only constraint defines the mass shell,
\be
p^2-m^2=0,
\label{wm}
\ee
or, in presence of the einbein field $\mu$,
\be
\pi=0,\quad H=-\frac{p^2-m^2}{2\mu},
\ee
with $\pi$ being the momentum canonically conjugated to $\mu$ and $H$ being the Hamiltonian
function of the
system (in case (\ref{wm}) it identically vanishes). The requirement that the constraint
$\pi=0$ be preserved in time returns one to the mass-shell condition (\ref{wm}):
\be
0=\dot{\pi}=\{\pi H\}=\frac{\partial H}{\partial \mu}=\frac{p^2-m^2}{2\mu^2}\sim p^2-m^2.
\ee

To make things simpler, one can fix the gauge-like freedom (\ref{gl}) identifying the
proper time $\tau$ with one of the physical coordinates of the particle. The most popular
choices are
\begin{itemize}
\item the laboratory gauge ($\tau=x_0$);
\item the proper time gauge ($\tau=(nx)$, $n_{\mu}=\frac{P_{\mu}}{\sqrt{P^2}}$ with
$P_{\mu}$ being the total momentum of the system) \cite{Rh};
\item the light-cone gauge ($\tau=\frac12(x_0+x_3)=x_+$),
\end{itemize}
which lead to quantization of the system on different hypersurfaces.

With the laboratory gauge fixed the Lagrangian function (\ref{L1}) becomes
\be
L=-\frac{m^2}{2\mu}-\frac{\mu}{2}+\frac{\mu\dot{\vec{x}}^2}{2},
\label{LLL}
\ee
so that the corresponding Hamiltonian function reads
\be
H=\frac{\vec{p}^2+m^2}{2\mu}+\frac{\mu}{2},
\label{H1}
\ee
and after taking the extremum in $\mu$ one ends with the standard relativistic expression
\be
H=\sqrt{\vec{p}^2+m^2}.
\ee

\subsection{Einbeins as variational parameters}\label{subsec22}

In the simple example considered above neither the Lagrange nor the Hamilton functions of the system
contained $\dot{\mu}$, which allowed one to get rid of $\mu$ at any stage by taking the extremum
in the latter. It is not so for more complicated systems when a change of variables is to
be performed which touches upon the einbeins. The velocity corresponding to the original
degrees of freedom of the system may mix in a very tangled way with those for einbeins, so
that it is not a simple task anymore to follow the lines {\it $\acute a$ la} Dirac \cite{Dirac} to resolve
the set of constraints and to get rid of
nonphysical degrees of freedom. See, $e.g.$, \cite{we-ein,spin,we-ein2} for several examples when such a
resolution can be done explicitly.

Luckily another approach to einbeins is known \cite{DKS,yuaein}. They can be treated as variational
parameters. Thus one replaces the dynamical function of time $\mu(\tau)$ by the
parameter $\mu_0$ independent of $\tau$. The eigenstate problem is solved then, keeping
$\mu_0$ constant, so that one has the spectrum $M_{\{n\}}(\mu_0)$, where $\{n\}$ denotes
the full set of quantum numbers. Then one is to minimize each eigenenergy independently with
respect to $\mu_0$\footnote{Note that solutions for $\mu_0$ of both signs appear, but only one of them
($\mu_0>0$) is finally left. Neglecting the negative solution is the general lack of the einbein
field approach and this leads to the fact that quark Zitterbewegung is not taken into
account (see also the discussion in Subsection \ref{subsec34}).}\label{fff}:
\be
\left.\frac{\partial M_{\{n\}}(\mu_0)}{\partial\mu_0}\right|_{\mu_0=\mu_0^*}=0,\quad\quad
M_{\{n\}}=M_{\{n\}}(\mu_0^*).
\ee

Such an approach has a number of advantages. First, it allows one to avoid the tedious algebra
of commuting constraints with one another following the
standard Dirac technique \cite{Dirac}. Second,
it allows one a very simple and physically transparent interpretation of einbeins. Indeed, in
formulae (\ref{L1}) and (\ref{LLL}) the einbein $\mu$ can be treated as an effective mass of the
particle; the dynamics of the system remains essentially relativistic, though being
nonrelativistic in form. If $m$ is the current quark mass, then $\mu$ can
be viewed as its constituent mass celebrated in hadronic phenomenology. What is more, the
current mass can be even put to zero, whereas the Lagrangian approach remains valid
in the presence of the einbeins and the standard Hamiltonian technique can be
developed then. The latter observation is intensively used in analytic QCD calculations for
glue describing gluonic degrees of freedom in glueballs and hybrids \cite{hybrid,glbl}.

An obvious disadvantage of the variational approach to the einbein fields is some loss of
accuracy. As a variational method it provides only an approximate solution, giving no hint
as to how to estimate the ultimate accuracy of the results. Thus in the next section we test
this method, comparing its predictions with exact solutions of some relativistic equations.
We consider the accuracy, found to be about 7\% at worst, quite reasonable, which justifies
our consequent attack on the light-light and heavy-light mesons spectra using this
formalism.

\section{Testing the method}\label{sec3}

\subsection{Quasiclassics for the spinless Salpeter equation}\label{subsec31}

We start from the Salpeter equation for the quark-antiquark system with equal masses and
restrict ourselves to the zero-angular-momentum case for simplicity:
\be
\left(2\sqrt{p_r^2+m^2}+\sigma r\right) \psi_n=M^{(ll)}_n\psi_n,
\label{ll}
\ee
where the subscript $(ll)$ stands for the light-light system.

The quasiclassical quantization condition looks like
\be
\int^{r_+}_0 p_r(r) dr=\pi\left(n+\frac{3}{4}\right),\quad
n=0,1,2,\ldots,\quad r_+=\frac{M^{(ll)}_n-2m}{\sigma},
\label{8}
\ee
where the integral on the l.h.s. can be worked out analytically, yielding
\be
M^{(ll)}_n\sqrt{\left(M^{(ll)}_n\right)^2-4m^2}-4m^2\ln
\frac{\sqrt{\left(M_n^{(ll)}\right)^2-4m^2}+M^{(ll)}_n}{2m}=4\sigma
\pi \left(n+\frac{3}{4}\right),
\label{bsll}
\ee
or approximately $(m\ll\sqrt{\sigma})$ one has
\be
\left(M^{(ll)}\right)^2=4\pi\sigma\left(n+\frac34\right)+2m^2\ln\frac{\pi\sigma(n+3/4)}{m^2}+\ldots .
\label{st}
\ee

Solution (\ref{st}) becomes exact in the limit $m=0$, whereas for a nonzero mass the
leading
correction to the linear regime $(M^{(ll)})^2\sim n$ behaves like
\be
\frac{\Delta M^2_n}{M_n^2}=O\left(\frac{m^2}{M_n^2}\ln\frac{M_n}{m}\right)\lmn\frac{\ln n}{n}.
\label{Mnas}
\ee

For a heavy-light system one has the Salpeter equation
\be
\left(\sqrt{p_r^2+m^2}+\sigma r\right)\psi_n=M_n^{(hl)}\psi_n,
\label{hl}
\ee
where $M_n^{(hl)}$ denotes the excess over the heavy particle mass $M$.
Similarly to (\ref{bsll}) one finds then
\be
M_n^{(hl)}\sqrt{\left(M^{(hl)}_n\right)^2-m^2}-m^2\ln
\frac{\sqrt{\left(M^{(hl)}_n\right)^2-m^2}+M_n^{(hl)}}{m}=2\sigma \pi
\left(n+\frac{3}{4}\right),
\label{bshl}
\ee
and formula (\ref{Mnas}) holds true in this case as well.

Comparing the results of the WKB method with the exact solutions of
equation (\ref{ll}) (rows $M_n({\rm WKB})$ and $M_n({\rm exact})$ in Table \ref{T4}), 
one can see that the error does not exceed 3-4\% even
for the ground state. See also \cite{wkb_new} where the WKB method is tested
for light-light mesons.

\subsection{Quasiclassics for the one-particle Dirac equation}\label{subsec32}

As a next example we discuss the one-particle Dirac equation with linearly rising
confining potential \cite{wkb1}:
\be
(\vec \alpha \vec p + \beta (m+U)+V)\psi_n= \varepsilon_n\psi_n.
\label{Dir}
\ee

The WKB method applied to this equation gives \cite{wkb2,yas}
\be
\int^{r_+}_{r_-}\left(p+\frac{\kappa
w}{pr}\right)dr=\pi\left(n+\frac{1}{2}\right),\quad n=0,1,2,\ldots,
\label{14}
\ee
where
\be
p=\sqrt{(\varepsilon-V)^2-\frac{\kappa^2}{r^2}-(m+U)^2},
\label{15}
\ee
$$
w=-\frac{1}{2r}
-\frac{1}{2}\frac{U'-V'}{m+U+\varepsilon-V},
$$
$$
|\kappa|=j+\frac{1}{2}.
$$

For the most interesting case of purely scalar confinement ($V=0$, $U=\sigma r$) an
approximate quasiclassical solution was found in \cite{yas} ($m=0$):
\be
\varepsilon^2_n=2\sigma\left(2n+j+\frac{3}{2}+\frac{sgn\kappa}{2}
+\frac{\kappa\sigma}{\pi\varepsilon^2_n}\left(0.38+ln
\frac{\varepsilon^2_n}{\sigma|\kappa|}\right)+
 O\left(\left(\frac{\kappa\sigma}{\varepsilon_n^2}\right)^2\right)\right).
\label{WKBSim}
\ee

A detailed comparison of the results of the WKB method and those following from the 
recursive
formula (\ref{WKBSim}) with exact numerical solutions to equation (\ref{Dir}) is given
in \cite{MNS}. Here we only note that the
coincidence of the three numbers is impressive as even for the lowest states the
discrepancy does not exceed 1\%.

\subsection{Quasiclassical variational einbein field (combined) method for the spinless Salpeter
equ\-ation}\label{subsec33}

Finally we combine the two methods discussed above and apply the WKB approximation to the
Hamiltonian of a relativistic system with einbeins introduced as variational
parameters. Then the resulting quasiclassical spectrum is minimized with respect to the
einbeins. Thus we have a powerful method of solving the eigenvalues problem for
various relativistic systems which we call \lq\lq combined." Let us test the accuracy of
this method first.

We start from the Salpeter equation (\ref{ll}) for the light-light system and introduce
the parameter $\mu_0$ as described in Section \ref{sec2}:
\be
H_{1}=2\sqrt{p_r^2+m^2}+\sigma r\longrightarrow H_{2}=\frac{p_r^2+m^2}{\mu_0}+\mu_0+\sigma
r.
\label{H1toH2}
\ee
In what follows we consider the massless case substituting $m=0$ into (\ref{H1toH2}).

We give the analytic formulae for the spectrum of the Salpeter equation (\ref{ll})
obtained using the quasiclassical approximation for the Hamiltonian $H_1$
(following from equation (\ref{bsll}) for $m=0$), the exact solution for the
Hamiltonian $H_2$ minimized with respect to the einbein field and the result of
the combined method when the Bohr-Sommerfeld quantization condition is applied to the
Hamiltonian $H_2$ and the ultimate spectrum is also minimized with respect to $\mu_0$.
\be
M_n^2({\rm WKB})=4\pi\sigma\left(n+\frac34\right),
\label{W}
\ee
\be
M_n^2({\rm einbein})=16\sigma\left(\frac{-\zeta_{n+1}}{3}\right)^{3/2},
\label{ein}
\ee
\be
M_n^2({\rm combined})=\frac{8\pi}{\sqrt{3}}\sigma\left(n+\frac34\right),
\label{com}
\ee
where $\zeta_{n+1}$ is the $(n+1)$th zero of the Airy function $Ai(z)$ and counting of
zeros starts from unity. The extremal values of the einbein field in the latter two
cases read
\be
\mu_0^*({\rm einbein})=\sqrt{\sigma}\left(\frac{-\zeta_{n+1}}{3}\right)^{3/4},\quad
\mu_0^*({\rm combined})=\sqrt{\frac{\sigma (n+3/4)}{2\sqrt{3}}},
\ee
$i.e.$, the effective quark mass is $\mu_0^*\sim\sqrt{\sigma}$ and it appears entirely due to the
interquark interaction.

In Table~\ref{T4} we compare the results of the above three approximate methods of solving
the eigenvalues problem for equation (\ref{ll}) with the exact solution. In the last row
we give the accuracy of the combined method {\it vs} the exact solution. Two
conclusions can be deduced from Table~\ref{T4}. The first one is that the accuracy of all
approximate methods is high enough, including the combined method, which is of most interest
for us in view of its consequent applications to the QCD string with quarks at the ends.
The other conclusion is that the variational einbein field method gives a systematic
overestimation for the excited states which is of order 5-7\%.

\subsection{Discussion}\label{subsec34}

Here we would like to make a couple of concluding comments concerning the numerical
methods tested in this section, their accuracy and possible ways of their improvement.
As stated above the combined quasiclassical variational method is of most interest for us,
so we shall concentrate basically on it. The following two remarks are in order here.

From Table~\ref{T4} one can see that the relative error is practically constant, tending to
the value of 7\% for large $n$. The reason for such a behaviour will become clear
if one compares
formulae (\ref{W}) and (\ref{com}). Both relations reproduce the same dependence on the
radial quantum number $n$, whereas the difference comes from different slopes, $4\pi\sigma$
in (\ref{W}) $vs$ $\frac{8\pi}{\sqrt{3}}\sigma$ in (\ref{com}). Then for highly excited
states the error is practically independent of $n$ and can be estimated as
\be
\delta=\frac{M_n({\rm combined})-M_n({\rm exact})}{M_n({\rm combined})}\approx
\frac{M_n({\rm combined})-M_n({\rm WKB})}{M_n({\rm combined})}=
1-\sqrt{\frac{\sqrt{3}}{2}}\approx 0.07;
\ee
$i.e.$, the ultimate accuracy of the quasiclassical variational einbein field method
(combined method) appears to be about 7\%.
Introducing, say, a correcting factor in (\ref{com}) one could overcome the systematic
overestimation and reproduce the spectrum with
a better accuracy. We shall return to this
observation later on when discussing the spectrum of the heavy-light mesons.

Another source of error in the einbein field approach is neglecting the quark
Zitterbewegung (see the footnote on page \pageref{fff}). As stated above we neglect the
negative sign solution for the einbein field $\mu_0$ expecting its small influence on
the spectrum. Let us give some reasoning to justify this action.

It was demonstrated numerically
in \cite{dM} that the contribution of the quark Z-graphs into $M^2$
is nearly constant for large excitation numbers and is of order 10\%, so
that the corresponding shift of $M$ behaves like
\be
\Delta M\sim\frac{\Delta M^2}{M}\lmn \frac{1}{\sqrt{n}},
\ee
so it is somewhat suppressed.

Besides, the good agreement of our numerical results with those provided by the
lattice data and taken from the Particle Data Group can also serve as an {\it a posteriori}
justification of such a neglect. Still some improvements for the einbein field approach
are needed to take this effect into account.

\section{Regge trajectories for relativistic equations with local potentials}\label{sec4}

It was observed long ago that the mesonic Regge trajectories are almost linear if the
total momentum or the radial quantum number is plotted {\it vs} the mesonic mass
squared~\cite{regge}:
\be
M^2(n,J)=c_nn+c_JJ+\Delta M^2,
\label{Rt}
\ee
where $c_n$ and $c_J$ are the (inverse) slopes while $\Delta M^2$ denotes corrections to the
leading linear regime which come from the self-energy, spin splittings, $etc$.

Relations like (\ref{Rt}) naturally appear in most of models for confinement, though the
(inverse) slopes $c_n$ and $c_J$ are different for different models.

For example the Salpeter equation for the heavy-light system,
\be
\left(\sqrt{p_r^2+p^2_{\theta,\varphi}+m^2}+\sigma r\right) \psi_{nl}=M_{nl}\psi_{nl},
\ee
gives
\be
c_J^{(hl)}(Salpeter)=4\sigma,\quad c_n^{(hl)}(Salpeter)=2\sigma,
\label{cch}
\ee
where the total momentum $J$ coincides with the orbital one
$l$.

For the light-light system one easily finds, from (\ref{cch}) by a trivial parameters
rescaling,
\be
c_J^{(ll)}(Salpeter)=8\sigma,\quad c_n^{(ll)}(Salpeter)=4\sigma.
\label{ccl}
\ee

The one-particle Dirac equation (\ref{Dir}) yields different slopes for different natures of
the confining force. Thus for the purely vector confinement (potential added to the energy
term\footnote{We disregard the problem of the Klein paradox here.}) one finds
\be
c_{J\;{\rm vec}}^{(hl)}(Dirac)=4\sigma,
\ee
whereas, for purely scalar confinement (potential added to the mass term),
\be
c_{J\;{\rm scal}}^{(hl)}(Dirac)=2\sigma.
\ee

In the meantime the spectrum (\ref{Rt}) is expected to follow from a string-like picture
of confinement which predicts the (inverse) Regge slopes to be
\be
c_J^{(ll)}(string)=2\pi\sigma,\quad c_J^{(hl)}(string)=\pi\sigma.
\label{str}
\ee

One can easily see that none of the relativistic equations considered before gives the
correct result (\ref{str}); moreover the discrepancies are rather large (of order
25\%). See also \cite{Allen} for a discussion of various models of confinement and the
corresponding Regge trajectory slopes.

The reason why relativistic Salpeter and Dirac equations fail to reproduce the correct
string slope of the Regge trajectories is obvious and quite physically transparent.
Indeed, all relativistic equations with local potentials have only a trivial dependence of
the interquark interaction on the angular momentum which comes entirely from the quark
kinetic energy. Meanwhile, QCD is believed to lead to a string-type interaction between
the colour constituents inside hadrons, whereas the QCD string developed between quarks possesses its own
inertia and thus it should also contribute to the $J$-dependent part of the interaction.
It is this extra purely string-type piece of the interquark interaction to give an extra
contribution to the Regge trajectory slope and to bring it into the
correct form of (\ref{str}). This statement is proved explicitly in the next section,
whereas the string dynamics footprint in the heavy-light mesons spectrum is discussed in
detail in Section \ref{sec7}.

\section{Hamiltonian of the $q \bar q$ meson}\label{sec5}

\subsection{Quark-antiquark Green's function}\label{subsec51}

We start from the Euclidean Green's function of the $q \bar q$ pair in the
confining vacuum
\be
G_{q\bar q}=\langle\Psi_{q\bar q}^{(f)}(\bar{x},\bar{y}|A)^+\Psi^{(i)}_{q\bar
q}(x,y|A)\rangle_{q\bar{q}A},
\label{Gqq}
\ee
where the initial and the final mesonic wave functions
\be
\Psi^{(i,f)}_{q\bar q}(x,y|A)=\bar{\Psi}_{\bar q}(x)\Phi(x,y)\Gamma^{(i,f)}\Psi_q(y)
\ee
are gauge invariant due to the standard path-ordered parallel transporter
\be
\Phi(x,y)=P\exp{\left(ig\int_{y}^{x}dz_{\mu}A_{\mu}\right)},
\ee
$\Gamma^{(i,f)}$ denote the matrices which might be inserted into the initial and final
meson-quark-antiquark vertices.

Integrating out the quark fields in (\ref{Gqq}), one finds, for the mesonic Green's
function,
\be
G_{q\bar q}=\langle Tr\Gamma^{(f)}S_q(\bar{x},x|A)\Phi(x,y)\Gamma^{(i)}S_{\bar{q}}
(y,\bar{y}|A)\Phi(\bar{y},\bar{x})\rangle_A,
\ee
where the trace stands for both colour and spinor indices. We have neglected
here the $1/N_C$-suppressed quark determinant, describing sea quark pairs.

To proceed further we employ the Feynman-Schwinger representation for the one-fermion
propagators in the external field, fix the laboratory gauge for both particles,
\be
x_{10}=t_1,\quad x_{20}=t_2,
\ee
and introduce the einbein fields $\mu_1$ and $\mu_2$ by means of the following change of variables
(see \cite{Green,DKS} for details):
\be
\mu_i(t_i)=\frac{T}{2s_i}\dot{x}_{i0}(t_i),\quad ds_iDx_{i0}\to D\mu_i(t_i),\quad i=1,2,
\ee
where $s_{1,2}$ are the Schwinger times, $T=\frac12(x_0+\bar{x}_0-y_0-\bar{y}_0)$.

Then the resulting expression for the mesonic Green's function reads \cite{Green}
\be
G_{q\bar q}=\int D\mu_1(t_1)D\mu_2(t_2)D\vec{x}_1D\vec{x}_2 e^{-K_1-K_2}
Tr\left[\vphantom{\int_0^T\frac{d}{d\mu}}\Gamma^{(f)}(m_1-\hat D)
\Gamma^{(i)}(m_2-\hat D)\times\right.
\label{Gqqf}
\ee
$$
\left.
P_{\sigma}\exp{\left(\int^T_0
\frac{dt_1}{2\mu_1(t_1)}\sigma^{(1)}_{\mu\nu}\frac{\delta}{i\delta
s_{\mu\nu} (x_1(t_1))}\right)}\exp{\left(-\int^T_0\frac{d t_2}{2\mu_2(t_2)}
\sigma^{(2)}_{\mu\nu}\frac{\delta}{i\delta s_{\mu\nu}(x_2(t_2))}\right)}
\exp{\left(-\sigma S_{\rm min}\right)}\right],
$$
with $K_i$ being the kinetic energies of the quarks
\be
K_i=\int_0^Tdt_i\left(\frac{m_i^2}{2\mu_i}+\frac{\mu_i}{2}+
\frac{\mu_i\dot{\vec{x}}_i^2}{2}\right),\quad i=1,2,
\label{Ks}
\ee
$\sigma_{\mu\nu}=\frac{1}{4i}(\gamma_{\mu}\gamma_{\nu}-
\gamma_{\nu}\gamma_{\mu})$, and $\delta/\delta s_{\mu\nu}$ denotes the
derivative with respect to the element of the area $S$. We have also used the minimal area
law asymptotic for an isolated Wilson loop,
\be
\left\langle TrP\exp{\left(ig\oint_Cdz_{\mu}A_{\mu}\right)}\right\rangle_A\sim \exp{(-\sigma S_{\rm
min})},
\ee
which is usually assumed for the stochastic QCD vacuum (see, $e.g.$, \cite{VCM}) and
found on the lattice. Here $S_{\rm min}$ is the area
of the minimal surface swept by the quark and antiquark trajectories.

Looking at (\ref{Gqqf}) one can easily recognize the following three main ingredients:
the contribution of the quark, the one of the antiquark, and finally the confining
interaction given by the string with tension $\sigma$. One can write, for the latter,
\be
S_{\rm min}=\int_0^Tdt\int_0^1d\beta\sqrt{(\dot{w}w')^2-\dot{w}^2w'^2},
\ee
with $w_{\mu}(t,\beta)$ being the string profile function chosen in linear form,
\be
w_{\mu}(t,\beta)=\beta x_{1\mu}(t)+(1-\beta)x_{2\mu},
\label{spr}
\ee
thus describing the straight-line string which is a reasonable approximation for the
minimal surface \cite{DKS}.

Finally, synchronizing the quark and the antiquark times $(t_1=t_2=t)$
one finds from (\ref{Gqqf}) that in the spinless approximation the quark-antiquark
meson can be described by the Lagrangian
\be
L(t)=-\frac{m_1^2}{2\mu_1}-\frac{m_2^2}{2\mu_2}-\frac{\mu_1+\mu_2}{2}+
\frac{\mu_1\dot{\vec{x}}_1^2}{2}+\frac{\mu_2\dot{\vec{x}}_2^2}{2}
-\sigma
r\int_0^1d\beta\sqrt{1-[\vec{n}\times(\beta\dot{\vec{x}}_1+(1-\beta)\dot{\vec{x}}_2)]^2},
\label{Lm}
\ee
where $\vec{r}=\vec{x}_1-\vec{x}_2$ and $\vec{n}=\vec{r}/r$.
Expansion of the surface-ordered exponents in (\ref{Gqqf}) gives a set of
spin-dependent corrections to the leading regime (\ref{Lm}).

\subsection{Hamiltonian for spinless quarks}\label{subsec52}

Starting from the Lagrangian (\ref{Lm}) and introducing an extra einbein field $\nu(t,\beta)$
continuously depending on the internal string coordinate $\beta$ one can get rid of the
square root in (\ref{Lm}) arriving at the Hamiltonian of the $q\bar q$ system in the
centre-of-mass frame in the form \cite{DKS}
$$
H=\sum_{i=1}^2\left(\frac{p_r^2+m_i^2}
{2\mu_i}+\frac{\mu_i}{2}\right)+\int^1_0d\beta\left(\frac{\sigma^2r^2}{2\nu}+
\frac{\nu}{2}\right)+\frac{\vec{L}^2}{2r^2[\mu_1{(1-\zeta)}^2+\mu_2{\zeta}^2+
\int^1_0d\beta\nu{(\beta-\zeta)}^2]},
$$
\be
\zeta=\frac{\mu_1+\int^1_0d\beta\nu\beta}{\mu_1+\mu_2+\int^1_0d\beta\nu}.
\label{Hm}
\ee

Similarly to $\mu$'s which have the meaning of the constituent quark masses, the einbein
$\nu$ can be viewed as the density of the string energy. In the simplest case of $l=0$ one
easily finds, for the extremal value of $\nu$,
\be
\nu_0=\sigma r;
\ee
$i.e.$, the energy distribution is uniform and the resulting interquark interaction is just
the linearly rising potential $\sigma r$. In the meantime, if $l\neq 0$, then
the two contributions can be identified in the last $l$-dependent term in (\ref{Hm}). Roughly speaking
the first two $\mu$-dependent terms in the denominator come from the
quark kinetic energy. The last term containing the integral over $\beta$ is nothing but
the extra inertia of the string discussed before. Rotating string also contributes to the
interquark interaction making it essentially nonlocal, so that the very notion of the
interquark potential is not applicable to the system anymore.

Note that the Hamiltonian (\ref{Hm}) has the form of sum of the \lq\lq kinetic" and the 
\lq\lq potential"
parts, but this is somewhat misleading, as extrema in all three einbeins are understood, so that
the ultimate form of the Hamiltonian would be extremely complicated and hardly available
for further analytical studies.

Expression (\ref{Hm}) can be simplified if one expands the Hamiltonian in powers of
$\sqrt{\sigma}/\mu$. One finds then \cite{DKS,KNn}
\be
H=H_0+V_{string},
\label{Hdec}
\ee
\be
H_0=\sum_{i=1}^2\left(\frac{\vec{p}^2+m_i^2}{2\mu_i}+\frac{\mu_i}{2}\right)+\sigma r,
\label{H0}
\ee
\be
V_{string}\approx -\frac{\sigma
(\mu_1^2+\mu_2^2-\mu_1\mu_2)}{6\mu_1^2\mu_2^2}\frac{\vec{L}^2}{r},
\label{Vstr}
\ee
where $V_{string}$ is known as the string correction \cite{MP,DKS} and this is the term totally missing
in the relativistic equations with local potentials.
Indeed, the Salpeter equation with the linearly rising potential is readily
reproduced from (\ref{H0}) if extrema in $\mu_{1,2}$ are taken explicitly, whereas
the string correction is lost. Meanwhile, its sign is
negative so that the contribution of the string lowers the energy of the system, thus
giving a negative contribution to the masses of orbitally excited states, leaving those with
$l=0$ intact. In Section \ref{sec6} we shall demonstrate how a proper account of the string
dynamics in the full Hamiltonian (\ref{Hm}) brings the Regge trajectory slope to the
correct value (\ref{str}), whereas in Section \ref{sec7} the string correction (\ref{Vstr}) will
be demonstrated to solve the problem of the identification of the resonance $D(2637)$
recently claimed by the DELPHI Collaboration \cite{DELPHI,DELPHI2}.

\subsection{Spin-dependent corrections}\label{subsec53}

Let us return to the quark-antiquark Green's function (\ref{Gqqf}) and extract the
nonperturbative spin-orbit interaction. Following \cite{Lisbon,BS} one finds
\be
V_{so}^{\rm np}=-\frac{\sigma}{2r}\left(\frac{\vec{S}_1\vec{L}}{\mu_1^2}+
\frac{\vec{S}_2\vec{L}}{\mu_2^2}\right).
\label{Vsd}
\ee

It follows from \cite{Lisbon,BS} that all potentials $V_i(r)$ (in the notation of
\cite{EFG})
contain both perturbative and nonperturbative pieces given there in explicit
form. One can argue that at large distances only the piece (\ref{Vsd}) is left
whereas for light quarks all nonperturbative ones may be important (see \cite{AM}).

Now, to have a full picture of the interquark interaction one is to supply
the purely nonperturbative string-type interaction described by the Hamiltonian (\ref{Hm})
by the perturbative gluon exchange adding the colour Coulomb potential to the Hamiltonian
$H_0$ from (\ref{H0}) and calculating the corresponding 
spin-dependent perturbative terms in addition to the potential 
(\ref{Vsd}). The result reads
\be
H_0=\sum_{i=1}^2\left(\frac{\vec{p}^2+m_i^2}{2\mu_i}+\frac{\mu_i}{2}\right)+\sigma
r-\frac43\frac{\alpha_s}{r}-C_0,
\label{H0new}
\ee
where we have also added the overall constant shift $C_0$ and
$$
V_{sd}=\frac{8\pi\kappa}{3\mu_1\mu_2}(\vec{S}_1\vec{S}_2)
\left|\psi(0)\right|^2-\frac{\sigma}{2r}\left(\frac{\vec{S}_1\vec{L}}{\mu_1^2}+
\frac{\vec{S}_2\vec{L}}{\mu_2^2}\right)
+\frac{\kappa}{r^3}\left(\frac{1}{2\mu_1}+\frac{1}{\mu_2}\right)
\frac{\vec{S}_1\vec{L}}{\mu_1}
+\frac{\kappa}{r^3}\left(\frac{1}{2\mu_2}+\frac{1}{\mu_1}\right)
\frac{\vec{S}_2\vec{L}}{\mu_2}
$$
\be
+\frac{\kappa}{\mu_1\mu_2r^3}\left(3(\vec{S}_1\vec{n})
(\vec{S}_2\vec{n})-(\vec{S}_1\vec{S}_2)\right)+\frac{\kappa^2}{2\pi\mu^2r^3}
\left(\vec{S}\vec{L}\right)(2-{\rm ln}(\mu r)-\gamma_E),\quad \gamma_E=0.57,
\label{Vsd2}
\ee
with $\kappa=\frac43\alpha_s$, $\mu=\frac{\mu_1\mu_2}{\mu_1+\mu_2}$. We have also
added the term of order $\alpha_s^2$
which comes from one-loop calculations and is intensively discussed in the
literature \cite{alpha2,gluelump,AM}. It is important to stress that $C_0$ is
due to the nonperturbative self-energy of light quarks, which explains the later
numerical inputs.

An important comment concerning the expansion (\ref{Vsd2}) is in order. Up to the last
term the expression (\ref{Vsd2}) coincides in form with the Eichten--Feiberg--Gromes
result \cite{EFG}, but we have effective quark masses $\mu_i$ in the denominators instead of
the current ones $m_i$. Once $\mu_i\sim\sqrt{\langle\vec{p}^2\rangle+m_i^2}>m_i$ or even
$\mu_i\gg m_i$, then the result (\ref{Vsd2}) is applicable to the case of light quark
flavours, when the expansion of the interaction in the inverse powers of the quark mass
$m_i$
obviously fails. The values of $\mu$'s are defined dynamically and differ from state to
state (see \cite{Lisbon} for details).

The Hamiltonian (\ref{H0new}) with spin-dependent terms (\ref{Vsd2}) will be used for
explicit calculations for heavy-light mesons. In the case of light-light states one
should include additional nonperturbative spin-dependent terms (see \cite{Lisbon} and
references therein). The masses of light-light mesons listed in Table~\ref{T5} have
been calculated from the Regge trajectories which do not take into account
spin-dependent terms and we give them for the sake of comparison. A more detailed
calculation for the light mesons taking these effects into account can be found in
\cite{AM}.

\section{Spectrum of light-light mesons}\label{sec6}

\subsection{Angular-momentum-dependent potential and Regge trajectories}\label{subsec61}

Starting from the Hamiltonian (\ref{Hm}) we stick with the case of equal
quark masses $m$
\be
H=\frac{p^2_r+m^2}{\mu}+\mu+\frac{{\vec L}^2/r^2}
{\mu+2\int^1_0(\beta-\frac{1}{2})^2\nu d\beta}
+\frac{\sigma^2 r^2}{2}\int^1_0\frac{d\beta}{\nu}+\int^1_0\frac{\nu}{2}d\beta.
\label{Hme}
\ee

The extremal value of the einbein field $\nu$ can be found explicitly and reads \cite{MNS}
\be
\nu_0(\beta)=\frac{\sigma r}{\sqrt{1-4y^2(\beta-\frac{1}{2})^2}},
\label{57}
\ee
where $y$ is the solution of the transcendental equation
\be
\frac{L}{\sigma r^2}=\frac{1}{4y^2}
(\arcsin~y-y\sqrt{1-y^2})+\frac{\mu{y}}{\sigma r},
\label{y}
\ee
and ${\vec L}^2=l(l+1)$.

For large angular momenta the contribution of the quarks (the last term on the r.h.s. of
(\ref{y})) is negligible so that the maximal possible value $y=1$ is reached, thus yielding
the solution for the free open string \cite{DKS} (see also the second entry in
\cite{we-ein2}).

With the extremal value $\nu_0$ from (\ref{57}) inserted, the Hamiltonian (\ref{Hme}) takes the form
\be
H=\frac{p^2_r+m^2}{\mu(\tau)}+\mu(\tau)+\frac{\sigma r}{y} \arcsin~y+\mu(\tau) y^2,
\label{Hwn}
\ee
with $y$ defined by equation (\ref{y}). The last two terms on the r.h.s. of equation
(\ref{Hwn}) can be considered as an effective \lq\lq potential"
\be
U(\mu,r)=\frac{\sigma r}{y} \arcsin~y+\mu(\tau) y^2,
\label{Ue}
\ee
which is nontrivially $l$ dependent. In Fig.1 we give the form of the effective potential
(\ref{Ue}) for a couple of low-lying states (solid line). It has the same asymptotic as the
naive sum of the linearly rising potential and the centrifugal barrier coming from the
kinetic energy of the quarks (dotted line). In the meantime it differs from the latter at finite
distances. The only exception is the case of zero angular momentum which should be treated
separately and leads to the linearly rising potential for any interquark separation.

\subsection{Numerical results}\label{subsec62}

Following the variational einbein field method described and tested above, we start from
the Hamiltonian (\ref{Hwn}) and change the einbein field $\mu$ for the variational
parameter $\mu_0$ \cite{MNS}, so that one has
\be
H=\frac{p^2_r+m^2}{\mu_0}+\mu_0+U(\mu_0,r),
\label{Hmu0}
\ee
\be
U(\mu_0,r)=\frac{\sigma r}{y}\arcsin~y+\mu_0 y^2.
\label{Umu0}
\ee

Then the quasiclassical method applied to the Hamiltonian (\ref{Hmu0}) gives
\be
\int^{r_+}_{r_-} p_r (r)dr=\pi\left(n+\frac{1}{2}\right),
\label{BZstr}
\ee
with
\be
p_r(r)=\sqrt{\mu_0(M-\mu_0-U(\mu_0,r))-m^2}.
\label{pr}
\ee

The eigenvalues $M_{nl}(\mu_0)$ for $m=0$ were found numerically from (\ref{BZstr}), (\ref{pr})
and the minimization procedure with respect to $\mu_0$ was used then.
Results for $M_{nl}$ are given in Table~\ref{T5}
and depicted in Fig.2 demonstrating very nearly straight lines with
approximately string slope $(2\pi\sigma)^{-1}$ in $l$ and twice as small slope
in $n$. Note that it is the region of intermediate values of $r$ that plays the crucial role
in the Bohr-Sommerfeld integral (\ref{BZstr}), $i.e.$, the region where the nontrivial
dependence of the effective potential $U(\mu_0,r)$ on the angular momentum is most
important (see Fig.1).

In Table~\ref{T55} we give a comparison of the masses of several light-light
mesonic states extracted by means of the numerical results from Table~\ref{T5}
with the experimental data and theoretical predictions taken from \cite{isgur2}.
We have fitted our results to the experimental spectrum using the negative
constant $\Delta M^2$ (see equation (\ref{Rt})).

\subsection{Discussion}\label{subsec63}

Let us recall the results obtained for the light-light mesons and discuss
problems connected to the given approach. The net result of the current section is the
$l$-dependent effective interquark potential which gives the naive linearly
rising interaction only for $l=0$. It was observed long ago \cite{numeric,DKS} that for large
angular momenta the quark dynamics is negligible and the slope (\ref{str}) naturally
appears from the picture of open rotating string. In the present paper we find that
for massless quarks even the low-lying mesonic states demonstrate nearly straight-line
Regge trajectories with string slope (\ref{str}).

One problem clearly seen from Figs.2,3 is the leading trajectory intercept
$l_0\equiv l\;(M^2=0)$. To reproduce the
experimental intercept around -0.5 (see Fig.3) starting from the theoretical one
+0.5 (see left plot in Fig.2) one needs
a large negative constant added either to the Hamiltonian (\ref{Hm}) (see,
$e.g.$, $C_0$ in equation
(\ref{H0new})) or in the form of $\Delta M^2$ directly in (\ref{Rt}) (see also
Table~\ref{T55}). Once the first way
might violate the linearity of the Regge trajectories, then one should expect QCD to prefer
the second one, though the first way remains more attractive from the practical point of
view and will be used in calculations of the heavy-light mesons spectrum in the next
section.

Another problem is that one of the most intriguing questions of mesonic
spectroscopy, the $\pi-\rho$ splitting (and a similar problem in the strange sector)
cannot be addressed in our model. Taking the exact solution of the spinless Salpeter
equation (\ref{ll}) with $n=l=0$ (see Table~\ref{T4} with an appropriate rescaling from
$\sigma=0.2GeV^2$ to $\sigma=0.17GeV^2$) one finds for the $\rho$ mass squared the value of order
$1.7GeV^2$ which does not violate the linearity of the trajectory (see the circled dot in
Fig.2). If the overall negative shift with $\sqrt{|\Delta M^2|}=1126MeV^2$ (see Table~\ref{T55})
is applied to this state, then one arrives at a $\rho$-meson mass about $775MeV$, $i.e.$,
a value very close to the experimental one. Note that we have practically coinciding constants for the
$\rho$- and $a$-meson trajectories (see the caption for Table~\ref{T55}), which supports
the idea that $\Delta M^2$ can be associated with quark self-energies.

\label{pionn}Meanwhile, one cannot pretend to describe pions (kaons) in the same framework as
their Goldstone nature is not implemented in the current model.
In realistic quantum-field-theory-based models each mesonic
state possesses two wave functions which describe the motion forward and backward in time
of the $q\bar q$ pair inside the meson \cite{twowave}. The backward motion is
suppressed if at least one of the quarks is heavy, for highly excited states and in
the infinite-momentum frame. For the chiral pion, which is expected to be strictly
massless in the chiral limit, the two wave functions are of the same order of
magnitude (see, $e.g.$, \cite{twowave2} for an explicit pionic solution in QCD$_2$), so
that none of them can be neglected. This explains why the naive estimate for the pion
mass lies much higher than the experimentally observed value of $140MeV$. For the
first excited state, $\rho$ meson, this effect is already suppressed, though one
still has to be careful to neglect the backward motion of the quarks. The progress in
this direction was achieved in recent papers by one of
the authors (Yu.S.) \cite{yas}, where a
Dirac-type equation was derived for the heavy-light system and the properties of its solutions
were investigated. This new formalism is expected to allow consideration of pionic Regge
trajectories as it has the chiral symmetry breaking built in.

\section{Spectrum of heavy-light mesons}\label{sec7}

All results obtained for the light-light mesons in Section \ref{sec6} can be reproduced for
the heavy-light states, so that in the one-body limit the Regge trajectories with the
correct string (inverse) slope $\pi\sigma$ are readily reproduced. Meanwhile, the aim of
this study is to take into account corrections to the leading regime which come from the
spin-dependent terms in the interquark interaction as well as those due to the finitness
of the heavy quark mass. Corrections of both types are important for establishing the
correct spectra of $D$ and $B$ mesons which are the main target of the present
investigation.

\subsection{Spectrum of the spinless heavy-light system}\label{subsec71}

In this subsection we study the spectrum of the heavy-light mesons, disregarding the quark
spins. This amounts to solving the Schr{\" o}dinger-like equation for the Hamiltonian $H_0$ from
(\ref{H0new}). Note that to this end one needs to know the nonrelativistic spectrum in
the potential which is the sum of the linearly rising and Coulomb parts
\cite{Lisbon,yuaDB,KNn}:
\be
\left(-\frac{d^2}{d\vec{x}^2}+|\vec{x}|-\frac{\lambda}{|\vec{x}|}\right)
\chi_{\lambda}=a(\lambda)\chi_{\lambda},
\label{Schr}
\ee
where
$$
\lambda=\kappa\left(\frac{2\mu}{\sqrt{\sigma}}\right)^{2/3},\quad\kappa=\frac43\alpha_s,\quad
\mu=\frac{\mu_1\mu_2}{\mu_1+\mu_2}.
$$

If solutions of (\ref{Schr}) for $\chi_{\lambda}$ and $a(\lambda)$ are known as functions
of the reduced Coulomb potential strength $\lambda$, then one can find the following
expressions for the extremal values of the einbeins (constituent quark masses):
\be
\mu_1(\lambda)=\sqrt{m_1^2+\Delta^2(\lambda)},\quad\mu_2(\lambda)=\sqrt{m_2^2+\Delta^2(\lambda)},
\quad\mu(\lambda)=\frac12\sqrt{\sigma}\left(\frac{\lambda}{\kappa}\right)^{3/2},
\label{mu1mu2}
\ee
with $\Delta(\lambda)$ given by
$$
\Delta^2(\lambda)=\frac{\sigma\lambda}{3\kappa}\left(a+2\lambda\left|\frac{\partial
a}{\partial\lambda}\right|\right).
$$

The definition of the reduced einbein field $\mu$ via $\mu_1$ and $\mu_2$ leads to the
equation defining~$\lambda$
\be
\mu(\lambda)=\frac{\mu_1(\lambda)\mu_2(\lambda)}{\mu_1(\lambda)+\mu_2(\lambda)}.
\label{lamb}
\ee

Technically this means that one should generate self-consistent solutions to equations
(\ref{Schr}) and (\ref{lamb}) which are subject to numerical calculations
\cite{yuaDB,KNn}. In Table~\ref{T6} we give such solutions for several radial and orbital
excitations in $D-$, $D_s-$, $B-$, and
$B_s-$mesonic spectra. We use the standard values for the string tension, the strong
coupling constant, and the current quarks masses. Note that $\alpha_s$ is chosen close to its frozen
value \cite{frozen} and it does not change a lot between $D$ and $B$ mesons. The reason is that in both
cases one has a light quark moving in the field of a very heavy one, so that the
one-gluon exchange depends on the size of the system, rather than on its total mass. Once
the difference in size between $D$ and $B$ mesons is not that large, the difference
between the two values of the strong coupling constant is also small (see Table~\ref{T6}).

The $\psi$ function at the origin given in the last
column of Table~\ref{T6} and which will be used later on for spin-spin splittings
is defined for radially excited states as
\be
\left|\psi(0)\right|^2=\frac{2\mu\sigma}{4\pi}\left(1+\lambda\langle
x^{-2}\rangle\right),
\label{psi0}
\ee
where
\be
\langle r^N\rangle=(2\mu\sigma)^{N/3}\langle x^N\rangle=(2\mu\sigma)^{N/3}\int_0^{\infty}x^{N+2}
\left|\chi_{\lambda}(x)\right|^2dx,\quad N>-3-2l,
\label{rav}
\ee
which immediately follows from the properties of equation (\ref{Schr}) and the
corresponding redefinition of variables.

\subsection{Spin-spin and spin-orbit splittings. The string correction}\label{subsec72}

In this subsection we calculate the spin-dependent corrections to the results given in
Table~\ref{T6} as well as those due to the proper string dynamics and which were intensively
discussed before.

The eigenstates of the Hamiltonian $H_0$ from (\ref{H0new}), which we consider to be the
zeroth approximation, can be specified in the form of terms $n^{2S+1}L_J$ ($n$ being the
radial quantum number) as the angular momentum $\vec{L}$, the total spin $\vec{S}$,
and the total momentum $\vec{J}=\vec{L}+\vec{S}$ are separately conserved by $H_0$.
The corresponding matrix elements for various operators present in (\ref{Vsd2}) read as
follows:
\be
\begin{array}{llll}
\hspace*{-1cm}^{2S+1}P_J\\
&\langle ^1P_1|\vec{S}_1\vec{L}|^1P_1\rangle=0,&
\langle ^1P_1|\vec{S}_2\vec{L}|^1P_1\rangle=0,&
\langle ^1P_1|(\vec{S}_1\vec{n})(\vec{S}_2\vec{n})|^1P_1\rangle=-\frac{\ds
1}{\ds 4},\\
{}\\
&\langle ^3P_0|\vec{S}_1\vec{L}|^3P_0\rangle=-1,&
\langle ^3P_0|\vec{S}_2\vec{L}|^3P_0\rangle=-1,&
\langle ^3P_0|(\vec{S}_1\vec{n})(\vec{S}_2\vec{n})|^3P_0\rangle=-\frac{\ds
1}{\ds 4},\\
{}\\
&\langle ^3P_1|\vec{S}_1\vec{L}|^3P_1\rangle=-\frac{\ds 1}{\ds 2},&
\langle ^3P_1|\vec{S}_2\vec{L}|^3P_1\rangle=-\frac{\ds 1}{\ds 2},&
\langle ^3P_1|(\vec{S}_1\vec{n})(\vec{S}_2\vec{n})|^3P_1\rangle=\frac{\ds
1}{\ds 4},\\
{}\\
&\langle ^3P_2|\vec{S}_1\vec{L}|^3P_2\rangle=\frac{\ds 1}{\ds 2},&
\langle ^3P_2|\vec{S}_2\vec{L}|^3P_2\rangle=\frac{\ds 1}{\ds 2},&
\langle ^3P_2|(\vec{S}_1\vec{n})(\vec{S}_2\vec{n})|^3P_2\rangle=\frac{\ds 1}{\ds 20};
\end{array}
\ee
\be
\begin{array}{llll}
\hspace*{-0.4cm}^{2S+1}D_J\\
&\langle ^1D_2|\vec{S}_1\vec{L}|^1D_2\rangle=0,&
\langle ^1D_2|\vec{S}_2\vec{L}|^1D_2\rangle=0,&
\langle ^1D_2|(\vec{S}_1\vec{n})(\vec{S}_2\vec{n})|^1D_2\rangle=-\frac{\ds
1}{\ds 4},\\
{}\\
&\langle ^3D_1|\vec{S}_1\vec{L}|^3D_1\rangle=-\frac{\ds 3}{\ds 2},&
\langle ^3D_1|\vec{S}_2\vec{L}|^3D_1\rangle=-\frac{\ds 3}{\ds 2},&
\langle ^3D_1|(\vec{S}_1\vec{n})(\vec{S}_2\vec{n})|^3D_1\rangle=-\frac{\ds
1}{\ds 12},\\
{}\\
&\langle ^3D_2|\vec{S}_1\vec{L}|^3D_2\rangle=-\frac{\ds 1}{\ds 2},&
\langle ^3D_2|\vec{S}_2\vec{L}|^3D_2\rangle=-\frac{\ds 1}{\ds 2},&
\langle ^3D_2|(\vec{S}_1\vec{n})(\vec{S}_2\vec{n})|^3D_2\rangle=\frac{\ds 1}{\ds 4},\\
{}\\
&\langle ^3D_3|\vec{S}_1\vec{L}|^3D_3\rangle=1,&
\langle ^3D_3|\vec{S}_2\vec{L}|^3D_3\rangle=1,&
\langle ^3D_3|(\vec{S}_1\vec{n})(\vec{S}_2\vec{n})|^3D_3\rangle=\frac{\ds
1}{\ds 28}.\\
\end{array}
\ee

The interaction $V_{sd}$ given by (\ref{Vsd2}) mixes orbitally excited states with
different spins, so that the transition matrix elements are given by
\be
\begin{array}{ll}
\langle ^1P_1|\vec{S}_1\vec{L}|^3P_1\rangle=\frac{\ds 1}{\ds \sqrt{2}},&
\langle ^1P_1|\vec{S}_2\vec{L}|^3P_1\rangle=-\frac{\ds 1}{\ds \sqrt{2}},\\
{}\\
\langle ^1D_2|\vec{S}_1\vec{L}|^3D_2\rangle=\sqrt{\frac{\ds 3}{\ds 2}},&
\langle ^1D_2|\vec{S}_2\vec{L}|^3D_2\rangle=-\sqrt{\frac{\ds 3}{\ds 2}},
\end{array}
\label{tran}
\ee
which lead to mixing within $|^1P_1\rangle$, $|^3P_1\rangle$
and $|^1D_2\rangle$, $|^3D_2\rangle$ pairs
so that the physical states are subject to matrix equations of the following
type:
\be
\left|
\begin{array}{cc}
E_1-E&V_{12}\\V^*_{12}&E_2-E
\end{array}
\right|=0.
\ee

Another important ingredient is the string correction given by (\ref{Vstr}) which leads to
an extra negative shift for orbitally excited states
\be
\delta M_l\approx -\frac{\sigma
(\mu_1^2+\mu_2^2-\mu_1\mu_2)}{6\mu_1^2\mu_2^2}l(l+1)\langle r^{-1}\rangle.
\label{dMl}
\ee

Thus the model is totally fixed and
the only remaining fitting parameter is the overall spectrum shift $C_0$ which finally
takes the following values:
\be
C_0(D)=212MeV,\quad C_0(D_s)=124MeV,\quad C_0(B)=203MeV,\quad C_0(B_s)=124MeV.
\label{C0}
\ee

Note that $C_0$ does not depend on the heavy quark 
($C_0(D)\approx C_0(B)$, $C_0(D_s)\approx C_0(B_s)$) and is completely defined by the
properties of the light one. For states with two light quarks one would have an overall
negative shift $2C_0$, which gives the contribution $-4C_0(M_n-C_0)$ to $M_n^2$.
In case of the $\rho$ meson this provides a negative constant of order $1GeV$, $i.e.$, the
right value needed to bring the theoretical intercept $l_0$ into the correct experimental one
(see $|\Delta M^2|$ in Table~\ref{T55}).

\subsection{Comparison with the experimental and lattice data. \lq\lq Mystery" of the $D(2637)$
state}\label{subsec73}

In Tables \ref{T7} and \ref{T8} we compare the results of our numerical calculations for the spectrum and
splittings with experimental and recent lattice data as well as with the theoretical
predictions from \cite{isgur2} and \cite{Faustov}. The underlined figures in Table~\ref{T7} are considered as the most probable
candidates for the experimentally observed values. One can see good agreement between
our theoretical predictions and the experimental values, as well as with the lattice
calculations \cite{lattice,lattice2}. To demonstrate the relevance of the corrections due to the heavy mass we
consider a simplified system containing one infinitely heavy particle and the light one
having its real mass. The best fits for the experimental spectra with the results for such
simplified systems are also given in Table~\ref{T7} in the column entitled $M_{hl}^{(0)}$ for
comparison. One can easily see that corrections in the inverse powers of the heavy mass
are strongly needed to reproduce the experimental spectrum with a reasonable accuracy and
to remove degeneracy of $S$ states.

Now we are in the position to resolve the \lq\lq mystery" of the $D(2637)$ state (and a similar
one in the $B$-meson spectrum). This state was claimed recently by the DELPHI Collaboration
\cite{DELPHI,DELPHI2}, but once its quantum numbers were not defined, then there was a problem of the
identification of this state. In most quark models (see also Table~\ref{T7}) the first radial excitation
$J^P=0^-$ lies approximately in the desired region of mass, but estimates of the width of such a state
lead to a confusion, as all such estimates give values much larger than the width of about
$15MeV$ reported by DELPHI. The only would-be way out of the problem is to identify this
narrow state with orbital excitations with $J^P$ being $2^-$ or $3^-$. In spite of the
fact that orbitally excited states are really narrower than the radially excited ones and
can have a width compatible with the experimental value, the following two
objections can be made \cite{discussion}:
i) quark models predict orbitally excited mesons to be at least $50MeV$ heavier than
needed, and
ii) a neighboring slightly more massive state should be observed as well.

It follows from Table~\ref{T7} that we can remove both objections mentioned above (see
also \cite{KNn}).
Indeed, one can easily see that orbitally excited states $2^-$ and $3^-$ lie even somewhat
lower than the radial excitation $0^-$. The reason for that is the negative string correction
(\ref{dMl}) for the orbitally excited states which comes from the proper dynamics of the
string. Besides that, the single $D$-wave $3^-$ state is an even more probable candidate
for the role of the observed $D(2637)$ resonance than the lightest one from the pair of
$2^-$ states, so that the problem of the \lq\lq missing state" is also avoided.

Note that our predictions $D(2654)$, $D(2663)$, and $D(2664)$ give larger masses
compared to the experimental one. This must be a reflection of the general lack of the
variational einbein field method ($\mu$ technique) discussed before, which gives slightly
overestimated values for the spectrum of excited states (see Table~\ref{T4} and the discussion in
subsection \ref{subsec34}).

\subsection{A bridge to the Heavy Quark Effective Theory}\label{subsect74}

In this subsection we discuss the correspondence between our model and the Heavy Quark
Effective Theory (HQET) approach widely discussed in the literature (see
\cite{hqet1,hqet2} and references therein). We use the standard parametrization for
the heavy-light meson mass,
\be
M_{hl}=m_Q+\bar\Lambda-\frac{1}{2m_Q}(\lambda_1+d_H\lambda_2)+O\left(\frac{1}{m_Q^2}\right),
\label{mH}
\ee
where $m_Q$ is the mass of the heavy constituent, and the coefficient $d_H$ describes the
hyperfine splitting,
\be
d_H=\left\{
\begin{array}{ll}
+3,&{\rm for}\;0^-\;{\rm states},\\
-1,&{\rm for}\;1^-\;{\rm states},
\end{array}
\right.
\ee
whereas $\bar\Lambda$, $\lambda_1$, and $\lambda_2$ are free parameters which are
subject to theoretical investigation. The parameter $\lambda_2$ is directly connected
to the splitting between $1^3S_1$ and $1^1S_0$ states and can be estimated from the
experimental $B$-mesonic spectrum to be
\be
\lambda_2=\frac14(M_{B^*}-M_B)\approx 0.12GeV^2.
\label{lam2}
\ee

From Table~\ref{T8} one can easily find our prediction for $\lambda_2$
\be
\lambda_2\approx 0.16GeV^2,
\ee
which being slightly overestimated is still in reasonable agreement with the
experimental value (\ref{lam2}).

In the meantime our model allows direct calculation
of the parameters $\bar\Lambda$, $\lambda_1$, and $\lambda_2$ based on the
Hamiltonian (\ref{H0new}). We apply the variational procedure described above to
the idealized system with $m_1\equiv m_Q\to\infty$ and $m_2\to 0$. This yields
\be
\bar{\Lambda}=\sqrt{\sigma}\left[\sqrt{\frac{\kappa}{\lambda_0}}a(\lambda_0)+
\sqrt{\frac{\lambda_0}{12\kappa}\left(a(\lambda_0)+2\lambda_0\left|\frac{\partial
a}{\partial\lambda_0}\right|\right)}\right],
\label{Lamb}
\ee
where $\kappa=\frac43\alpha_s$, and $a(\lambda)$ is the dimensionless eigenvalue
introduced in (\ref{Schr}) (see subsection \ref{subsec71}); $\lambda_0$ is a
solution to the equation
\be
\lambda^2=\frac43\kappa^2\left(a+2\lambda\left|\frac{\partial a}{\partial\lambda}\right|\right).
\label{lam0}
\ee

Then one can extract the coefficient $\lambda_2$ from the first term in equation (\ref{Vsd2}):
\be
\lambda_2=-\frac{4\pi\kappa}{3\mu_2}|\psi(0)|^2=-\frac23\sigma\kappa\left(1+\lambda_0\langle
x^{-2}\rangle\right).
\label{ll2}
\ee

An analytical formula for $\lambda_1$ is also available, but it is rather
bulky and we do not give it here. For $\alpha_s=0.39$ one can find the
numerical solution to equation (\ref{lam0}) to be $\lambda_0=1.175$. The corresponding
values for $\bar\Lambda$, $\lambda_1$, and $\lambda_2$ are given in the first column of
Table~\ref{T9} where they are compared with the results of other approaches.

Another way to estimate the discussed constants is to find the best fit of the form
\be
M_{fit}=m_Q+\bar\Lambda+C_0-\frac{\lambda_1}{2m_Q},
\ee
with $\bar\Lambda$ and $\lambda_1$ being the fitting parameters and $C_0=203MeV$
taken from (\ref{C0}), for eigenvalues of the Hamiltonian (\ref{H0new}) with $m_Q$
varied around the bottom quark mass $m_b=4.8GeV$ (we use the region
$4GeV<m_Q<6GeV$). The coefficient $\lambda_2$ can be found using formula (\ref{ll2}) with
$\lambda_0$ changed for the exact solution for $\lambda$ taken from Table~\ref{T6}.
Results are listed in the second column of Table~\ref{T9}.

One can see our figures to be in general agreement with those found in other
approaches among which we mention the QCD sum rules method \cite{sumrules}, the inclusive
semileptonic $B$-meson decays \cite{Bdec}, and the Dyson-Schwinger equation for the system of
a light quark and a static antiquark \cite{simtjon}. We find the parameter $\lambda_1$ to be
rather sensitive to the strong coupling constant $\alpha_s$. For example, for $\alpha_s=0.3$ one has $\lambda_1=-0.38GeV^2$
which should be confronted with the value $\lambda_1=-0.506GeV^2$ from the first column of
Table~\ref{T9} found for $\alpha_s=0.39$.

All our predictions for $\lambda_2$ exceed the value given by equation (\ref{lam2}). The
reason is the slightly
overestimated value of $\psi(0)$ given by the variational einbein field approach.

One should appreciate the advantage of the einbein field method which allows one to obtain relatively simple
analytical formulae for various parameters and to investigate their dependence on the
strong coupling constant $\alpha_s$ (the dependence on the only dimensional
parameter $\sigma$ is uniquely restored, giving $\bar\Lambda\sim\sqrt{\sigma}$ and
$\lambda_1\sim\lambda_2\sim\sigma$).

\section{Conclusions}\label{sec8}

In conclusion let us briefly recall the main results obtained in the present paper.

We use the model for the QCD string with quarks at the ends 
to calculate the spectra of light-light and
heavy-light mesons. There are two main points in which we differ from other approaches to
the same problem based on various relativistic Hamiltonians and equations with local
potentials. The first point is that we do not introduce the constituent mass by hand. On
the contrary, starting from the current mass we naturally arrive at the effective quark
masses which appear due to the interaction. Moreover, the resulting effective mass is
large enough even for the lightest quarks and lowest states in the spectrum, so that the
spin-dependent terms in the interquark interaction can be treated as perturbations in
most cases (except for pions and kaons) and thus
accounted for in this way.

The second advantage of the method is that the dynamics of the QCD string naturally
enters the game and it can be studied systematically. A proper account of this
dynamics allows one to resolve several problems of the mesonic spectroscopy; namely, one can
find that the rotating string lowers the masses of orbitally excited states, bringing
the (inverse) slope of Regge trajectories to their correct values ($2\pi\sigma$ for
light-light and $\pi\sigma$ for heavy-light states). In addition, this allows one to resolve the
problem of the identification of the $D(2637)$ state recently claimed by the DELPHI
Collaboration and which is known to lead to a contradiction between its small width,
incompatible with the decay modes for the radial excitation, and its mass lying
considerably lower than the values predicted by the quark models for orbitally excited
states. In the meantime, taking into account the negative string correction contributing
into the masses of orbitally excited mesons readily resolves this \lq\lq mystery" for
the $D(2637)$ state as well as for its counterpart in the spectrum of $B$ mesons.

For the heavy-light system we extract the constants $\bar\Lambda$, $\lambda_1$, and
$\lambda_2$ used in the framework of the Heavy Quark Effective Theory, for
which we derive analytical formulae. We
find our numerical results to be in agreement with those obtained from the experimental data
and calculations in other approaches like QCD sum rules, inclusive semileptonic $B$-meson
decays, and the relativistic Dyson-Schwinger equation for the $q\bar Q$ system.

We also conclude that the string-like interaction favoured by QCD invalidates the very
notion of any local interquark potential, still leaving room to the einbein field method for the
Hamiltonian approach to the bound states of quarks and gluons in QCD. Being rather
accurate, this method still requires improvements
to increase the accuracy and to have the full control over it.
\bigskip

The authors are grateful to A.M.Badalian, P.Bicudo, R.N.Faustov, A.B.Kaidalov,
V.S.Po\-pov and E.Ri\-bei\-ro
for useful discussions and valuable comments and to V.L.Morgunov and B.L.G.Bak\-ker for providing
numerical solutions to some equations. One of the authors (A.N.) would like to thank the
staff of the Centro de F\'\i sica das Interac\c c\~oes Fundamentais (CFIF-IST) for cordial
hospitality during his stay in Lisbon.

Financial support of RFFI grants 00-02-17836 and 00-15-96786,
INTAS-RFFI grant IR-97-232 and INTAS CALL 2000, project \# 110 
is gratefully acknowledged. One of the authors (A.N.) is also supported via RFFI grant
01-02-06273.

\newpage

\begin{table}[ht]
\begin{center}
\begin{tabular}{|c|c|c|c|c|c|c|}
\hline
$n$&0&1&2&3&4&5\\
\hline
$M_n$(WKB)&1.373 & 2.097& 2.629& 3.070&3.455&3.802\\
\hline
$M_n$(einbein)& 1.483& 2.256& 2.826& 3.300& 3.713& 4.085\\
\hline
$M_n$(combined)& 1.475& 2.254& 2.825& 3.299& 3.713& 4.085\\
\hline
$M_n$(exact)& 1.412& 2.106& 2.634& 3.073& 3.457& 3.803\\
\hline
$\frac{M_n({\rm combined})-M_n({\rm exact})}{M_n({\rm combined})}$, \%
& 4.27& 6.57& 6.76& 6.85& 6.89& 6.90\\
\hline
\end{tabular}
\end{center}
\caption{Comparison of the numerical results of the three approximate methods of solving
the eigenvalues problem for equation (\ref{ll}) for $m=0$, $\sigma=0.2~GeV^2$ and
$l=0$
given by equations (\ref{W})-(\ref{com}) with the exact eigenenergies of
the Hamiltonian $H_1$ from (\ref{H1toH2}).}\label{T4}
\begin{center}
\begin{tabular}{|c|c|c|c|c|c|}
\hline
\hspace*{0.3cm}$n$  $l$&1&2&3&4&5\\
\hline
0&1.719&2.029&2.287&2.516&2.725\\
\hline
1&2.362&2.611&2.829&3.025&3.208\\
\hline
2&2.850&3.069&3.264&3.442&3.608\\
\hline
3&3.259&3.460&3.639&3.803&3.955\\
\hline
4&3.619&3.806&3.973&4.127&4.268\\
\hline
5&3.944&4.120&4.276&4.423&4.554\\
\hline
\end{tabular}
\end{center}
\caption{Quasiclassical spectrum of the Hamiltonian (\ref{Hmu0}), (\ref{Umu0}) for $m=0$ and
$\sigma=0.17~GeV^2$ minimized with respect to the einbein $\mu_0$ (combined method).}\label{T5}
\end{table}
\begin{table}[h]
\begin{center}
\begin{tabular}{|c|c|c|c|c|c|c|}
\hline
{\rm Meson}&$^{2S+1}L_J$&$J^{PC}$&$M_{exp},\;MeV$&$M_{theor},\;MeV$&
$M_{theor}$\cite{isgur2}$,\;MeV$&${\rm
Error},\;\%$\\
\hline
\hline
$\rho$&$^3S_1$&$1^{--}$&770&775&770&0.6\\
\hline
$\rho_3$&$^3D_3$&$3^{--}$&1690&1688&1680&0.1\\
\hline
$\rho_5$&$^3G_5$&$5^{--}$&2330&2250&2300&3.4\\
\hline
\hline
$a_1$&$^3P_1$&$1^{++}$&1260&1346&1240&6.8\\
\hline
$a_3$&$^3F_3$&$3^{++}$&2070&2021&2050&2.4\\
\hline
\hline
$a_2$&$^3P_2$&$2^{++}$&1320&1316&1310&0.5\\
\hline
$a_4$&$^3F_4$&$4^{++}$&2040&2002&2010&1.9\\
\hline
$a_6$&$^3H_6$&$6^{++}$&2450&2491&-&1.7\\
\hline
\end{tabular}
\end{center}
\caption{Comparison of the masses of the light-light mesons lying on the lowest Regge
trajectories ($n=0$, $S=1$, $J=l+1$ for the $\rho$ and the $a_2$ trajectories; $n=0$, $S=1$,
$J=l$ for the $a_1$ trajectory), calculated for $\sigma=0.17GeV^2$ and the
overall negative shifts $\sqrt{|\Delta M^2|}=1126MeV$ for the $\rho$ trajectory,
$\sqrt{|\Delta M^2|}=1070MeV$ for the $a_1$ trajectory, and
$\sqrt{|\Delta M^2|}=1105MeV$ for the $a_2$ trajectory, with the experimental data
and with the theoretical
predictions taken from \protect\cite{isgur2}. See also the discussion concerning the pion and the
$\rho$-meson masses in subsection \ref{subsec63}.}\label{T55}
\end{table}
\begin{table}[t]
\begin{center}
\begin{tabular}{|c|c|c|c|c|c|c|c|c|c|c|c|c|}
\hline
$n$&$l$&meson&$m_1$&$m_2$&$\sigma$&$\alpha_s$&$\lambda$&$\mu_1$&$\mu_2$&$\mu$&$E_0$&$|\psi(0)|$\\
\hline
0&0&$D$&1.4&0.009&0.17&0.4&0.817&1.497&0.529&0.391&2.198&0.161\\
\cline{3-13}
&&$D_s$&1.4&0.17&0.17&0.4&0.847&1.501&0.569&0.412&2.224&0.167\\
\cline{3-13}
&&$B$&4.8&0.005&0.17&0.39&0.999&4.840&0.619&0.549&5.527&0.209\\
\cline{3-13}
&&$B_s$&4.8&0.17&0.17&0.39&1.035&4.842&0.658&0.579&5.550&0.219\\
\hline
\hline
0&1&$D$&1.4&0.009&0.17&0.4&0.869&1.522&0.597&0.428&2.640&0\\
\cline{3-13}
&&$D_s$&1.4&0.17&0.17&0.4&0.891&1.525&0.629&0.445&2.663&0\\
\cline{3-13}
&&$B$&4.8&0.005&0.17&0.39&1.052&4.847&0.675&0.593&5.949&0\\
\cline{3-13}
&&$B_s$&4.8&0.17&0.17&0.39&1.080&4.849&0.707&0.617&5.970&0\\
\hline
\hline
0&2&$D$&1.4&0.009&0.17&0.4&0.924&1.554&0.674&0.470&2.961&0\\
\cline{3-13}
&&$D_s$&1.4&0.17&0.17&0.4&0.942&1.557&0.702&0.484&2.982&0\\
\cline{3-13}
&&$B$&4.8&0.005&0.17&0.39&1.128&4.860&0.762&0.659&6.245&0\\
\cline{3-13}
&&$B_s$&4.8&0.17&0.17&0.39&1.151&4.861&0.789&0.679&6.263&0\\
\hline
\hline
1&0&$D$&1.4&0.009&0.17&0.4&0.929&1.557&0.682&0.474&2.848&0.162\\
\cline{3-13}
&&$D_s$&1.4&0.17&0.17&0.4&0.947&1.561&0.710&0.488&2.869&0.165\\
\cline{3-13}
&&$B$&4.8&0.005&0.17&0.39&1.142&4.863&0.779&0.671&6.131&0.207\\
\cline{3-13}
&&$B_s$&4.8&0.17&0.17&0.39&1.165&4.864&0.806&0.692&6.149&0.212\\
\hline
\end{tabular}
\end{center}
\caption{Solutions of equations (\ref{Schr})-(\ref{lamb}) for standard values of the
string tension $\sigma$, the strong coupling constant $\alpha_s$, and the current masses of
the quarks. $E_0$ is the mass of the corresponding state. All parameters are given in $GeV$ to the appropriate
powers.}\label{T6}
\end{table}
\begin{table}[t]
\begin{center}
\begin{tabular}{|c|c|c|c|c|c|c|c|c|c|}
\hline
&$n^{2S+1}L_J$&$J^P$&$M_{exp}$&$M_{theor}$&$M_{hl}^{(0)}$&$M_{theor}\;\cite{isgur2}$&$M_{theor}\;\cite{Faustov}$&$M_{lat}\;\cite{lattice}$&$M_{lat}\;\cite{lattice2}$\\
\hline
\hline
$D$&$1^1S_0$&$0^-$&1869&1876&2000&1880&1875&1884&1857\\
\hline
$D^*$&$1^3S_1$&$1^-$&2010&2022&2000&2040&2009&1994&1974\\
\hline
$D_1$&$\low{1^1P_1/^3P_1}$&$\low{1^+}$&$\low{2420}$&2354&2393&$\low{2440/2490}$&$\low{2414/2501}$&&$\low{2405/2414}$\\
\cline{5-6}
&&&&$\underline{2403}$&2407&&&&\\
\hline
&$1^3P_0$&$0^+$&&2280&2400&2400&2438&&2444\\
\hline
$D_2$&$1^3P_2$&$2^+$&2460&2432&2400&2500&2459&&2445\\
\hline
&$1^3D_3$&$3^-$&&\underline{2654}&2600&2830&&&\\
\cline{2-3}\cline{5-7}
${D^*}'$&$\low{1^1D_2/^3D_2}$&$2^-$&$\low{2637}$&$\underline{2663}$&2573&&&&\\
\cline{5-6}
&&&&2729&2600&&&&\\
\cline{2-3}\cline{5-8}
&$2^3S_1$&$0^-$&&\underline{2664}&2500&2640&2629&&\\
\hline
\hline
$D_s$&$^1S_0$&$0^-$&1968&1990&2100&1980&1981&1984&{\rm input}\\
\hline
$D_s^*$&$1^3S_1$&$1^-$&2112&2137&2100&2130&2111&2087&2072\\
\hline
$D_{1s}$&$\low{1^1P_1/^3P_1}$&$1^+$&$\low{2536}$&2471&2494&$\low{2530/2570}$&$\low{2515/2569}$&$\low{2494}$&$\low{2500/2511}$\\
\cline{5-6}
&&&&$\underline{2516}$&2506&&&&\\
\hline
&$1^3P_0$&$0^+$&&2395&2500&2480&2508&&2499\\
\hline
$D_{2s}$&$1^3P_2$&$2^+$&2573&2547&2500&2590&2560&2411&2554\\
\hline
\hline
$B$&$1^1S_0$&$0^-$&5279&5277&5200&5310&5285&5293&5277\\
\hline
$B^*$&$1^3S_1$&$1^-$&5325&5340&5200&5370&5324&5322&5302\\
\hline
$B_1$&$\low{1^1P_1/^3P_1}$&$\low{1^+}$&$\low{5732}$&5685&5592&&$\low{5719/5757}$&&$\low{5684/5730}$\\
\cline{5-6}
&&&&\underline{5719}&5608&&&&\\
\hline
&$1^3P_0$&$0^+$&&5655&5600&5760&5738&&5754\\
\hline
$B_2$&$1^3P_2$&$2^+$&5731&5820&5600&5800&5733&&5770\\
\hline
&$1^3D_3$&$3^-$&&$\underline{5955}$&5800&6110&&&\\
\cline{2-3}\cline{5-7}
${B^*}'$&$\low{1^1D_2/^3D_2}$&$2^-$&$\low{5860}$&$\underline{5953}$&5773&&&&\\
\cline{5-6}
&&&&6018&5827&&&&\\
\cline{2-3}\cline{5-9}
&$2^3S_1$&$0^-$&&$\underline{5940}$&5700&5930&5898&5890&\\
\hline
\hline
$B_s$&$1^1S_0$&$0^-$&5369&5377&5400&5390&5375&5383&{\rm input}\\
\hline
$B^*_s$&$1^3S_1$&$1^-$&5416&5442&5400&5450&5412&5401&5395\\
\hline
$B_{1s}$&$\low{1^1P_1/^3P_1}$&$1^+$&$\low{5853}$&5789&5793&$\low{5860/5860}$&$\low{5831/5859}$&$\low{5783}$&$\low{5794/5818}$\\
\cline{5-6}
&&&&$\underline{5819}$&5807&&&&\\
\hline
&$1^3P_0$&$0^+$&&5757&5800&5830&5841&&5820\\
\hline
$B_{2s}$&$1^3P_2$&$2^+$&&5834&5800&5880&5844&5848&5847\\
\hline
\end{tabular}
\end{center}
\caption{Masses of the $D$, $D_s$, $B$, and $B_s$ mesons in $MeV$. For the lattice results
we give only the central values extracted
from Figures 26, 27 and Tables XXVIII, XXIX of \protect\cite{lattice} and from
Table VIII of \protect\cite{lattice2}. We also compare out results with theoretical predictions taken
from \protect\cite{isgur2} and \protect\cite{Faustov}.
The symbols $1^1P_1/^3P_1$ and $1^1D_2/^3D_2$ are used to indicate that the
physical states
are mixtures of the $1^1P_1$ and $1^3P_1$ or $1^1D_2$ and
$1^3D_2$ states, respectively. Underlined figures give masses of the most
probable candidates for the experimentally observed resonances. The column $M_{hl}^{(0)}$
contains the best fit to the experimental spectrum for the system containing one
particle being infinitely heavy.}\label{T7}
\begin{center}
\begin{tabular}{|c|c|c|c|c|c|c|c|c|}
\hline
Splitting&$D_s$--$D$&$D_s^*$--$D^*$&$D^*$--$D$&$D_s^*$--$D_s$&$B_s$--$B$&$B_s^*$--$B^*$&$B^*$--$B$&$B_s^*$--$B_s$\\
\hline
Experiment&99&102&141&144&90&91&46&47\\
\hline
Theory&114&115&146&147&100&102&63&65\\
\hline
Theory \cite{isgur2}&100&90&160&150&80&80&60&60\\
\hline
Lattice \cite{lattice}&100&92&110&103&90&90&30&29\\
\hline
Lattice \cite{lattice2}&112&98&117&103&92&93&25&29\\
\hline
\end{tabular}
\end{center}
\caption{Splittings for the $D$, $D_s$, $B$, and $B_s$ mesons in $MeV$. Lattice
results are taken from Tables XXVIII, XXIX of \protect\cite{lattice} and Table VIII of
\protect\cite{lattice2}. We give also results of the theoretical papers
\protect\cite{isgur2}.}\label{T8}
\end{table}
\begin{table}[ht]
\begin{center}
\begin{tabular}{|c|c|c|c|c|c|}
\hline
&$m_1\to\infty$ $m_2\to 0$&$B$ mesons&Sum rules \cite{sumrules}&$B$ mesons decays \cite{Bdec}&
DS equation \cite{simtjon}\footnote{Note that slightly different values
for the string tension ($\sigma=0.18GeV^2$) and the strong coupling constant
($\alpha_s=0.3$) were used in this paper.}\\
\hline
$\bar\Lambda$, $GeV$&0.471&0.485&0.4 $\div$ 0.5&0.39 $\pm$ 0.11&0.493/0.288\\
\hline
$\lambda_1$, $GeV^2$&-0.506&-0.379&-0.52 $\pm$ 0.12&-0.19 $\pm$ 0.10&-\\
\hline
$\lambda_2$, $GeV^2$&0.21&0.17&0.12&0.12&-\\
\hline
\end{tabular}
\end{center}
\caption{Standard parameters used in HQET (see equation
(\ref{mH})). In the first column we give the values following
from formulae (\ref{Lamb}) and (\ref{ll2}) for $\bar\Lambda$
and $\lambda_2$ and the corresponding one for $\lambda_1$; the
second column contains the best fit of the form
$m_Q+\bar\Lambda+C_0-\frac{\lambda_1}{2m_Q}$ for the eigenvalues
of the Hamiltonian (\ref{H0new}) for $m_Q$ in the region
$4GeV<m_Q<6GeV$ (around $m_b=4.8GeV$), $C_0=203MeV$ being the
overall negative shift constant taken from equation (\ref{C0}). The
two numbers given in the last column correspond to local and nonlocal
kernels of the Dyson-Schwinger (DS) equation (see \protect\cite{simtjon} for
details). The figures given in the first two columns correspond
to $\sigma=0.17GeV^2$ and $\alpha_s=0.39$.}\label{T9}
\end{table}
\begin{figure}[ht]
\centerline{\epsfig{file=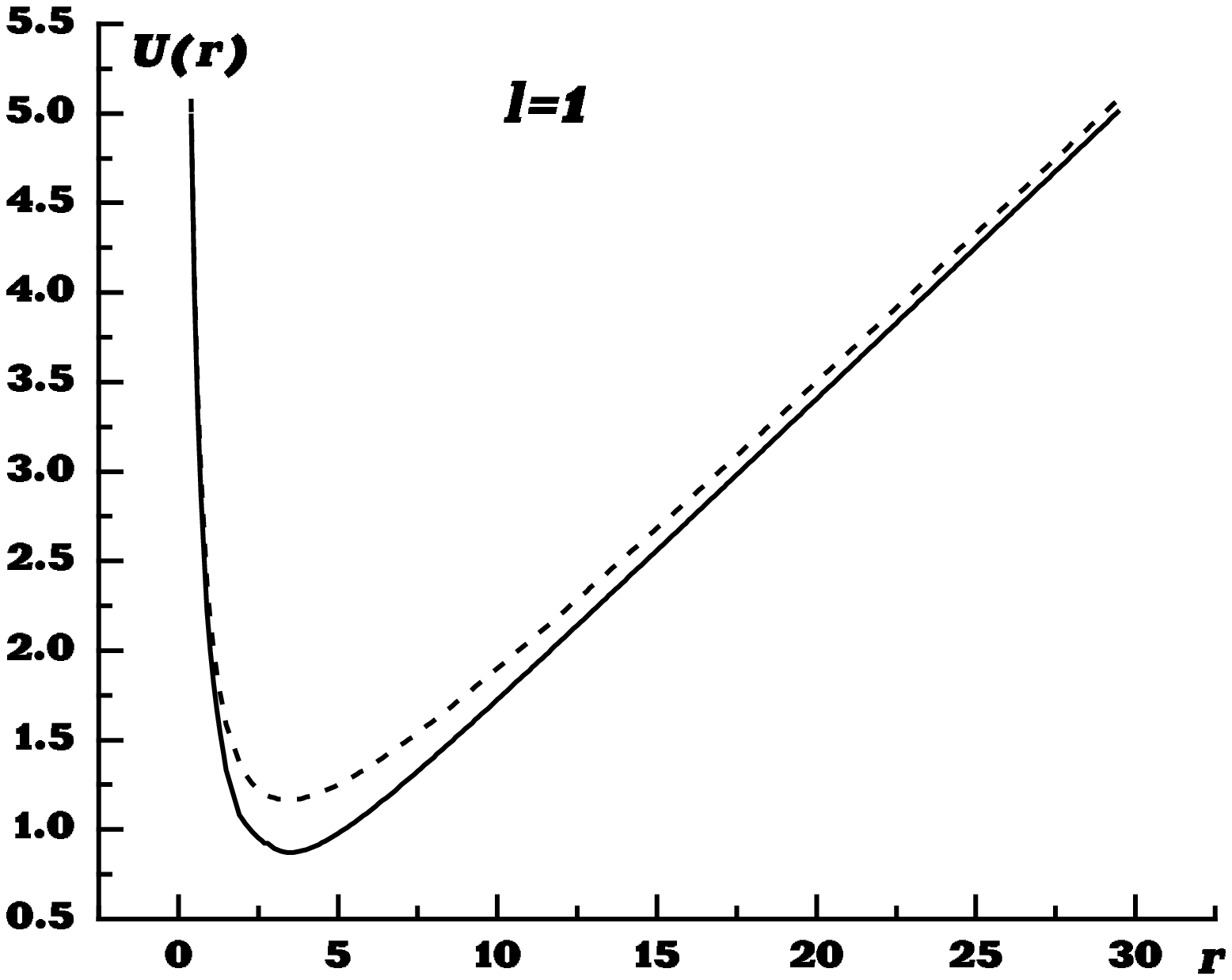,width=10cm}\hspace{-1.5cm}
            \epsfig{file=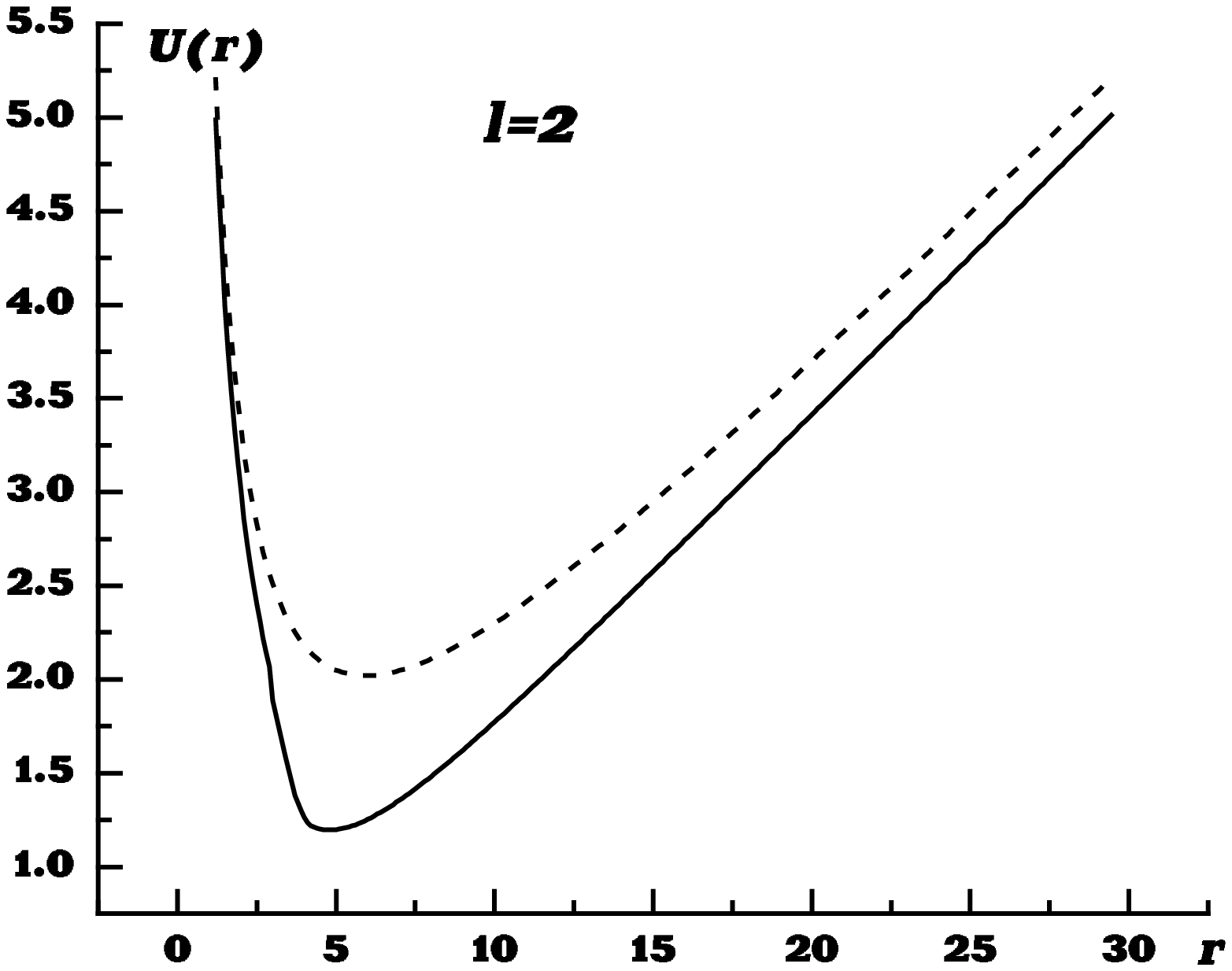,width=10cm}}
\caption{Effective potential incorporating the string rotation as well as the quark radial
motion for $\sigma=0.17~GeV^2$ and two values of the angular momentum $l$ (solid line). The naive sum
of the quark centrufugal barrier for the given $l$ and the linearly rising potential
$\sigma r$ is given in each graph by the dotted line.
For $l=0$ the effective potential coincides with $\sigma r$ for all values of $r$.}
\end{figure}

\begin{figure}[ht]
\centerline{\epsfig{file=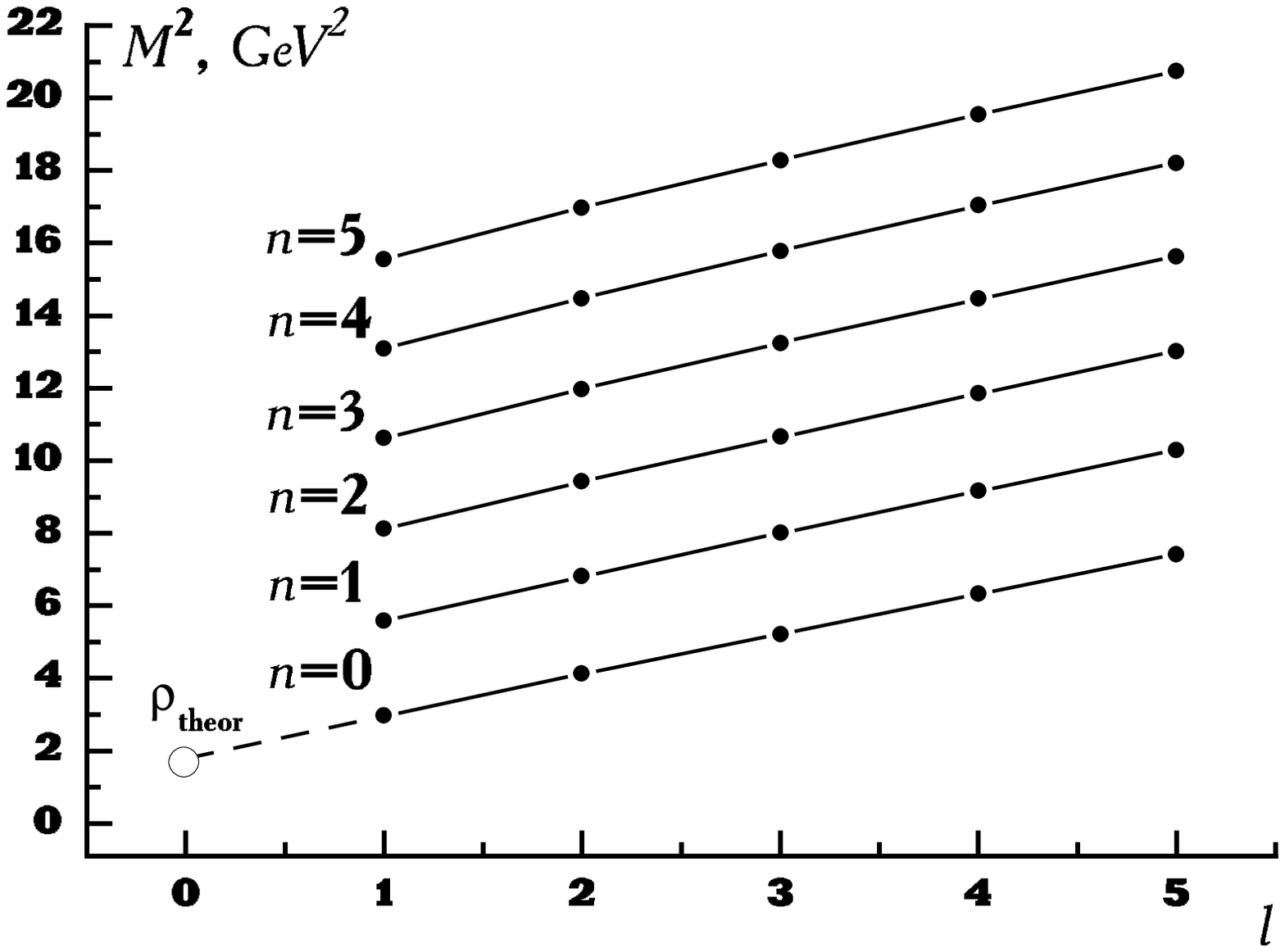,width=10cm}\hspace{-1.5cm}
            \epsfig{file=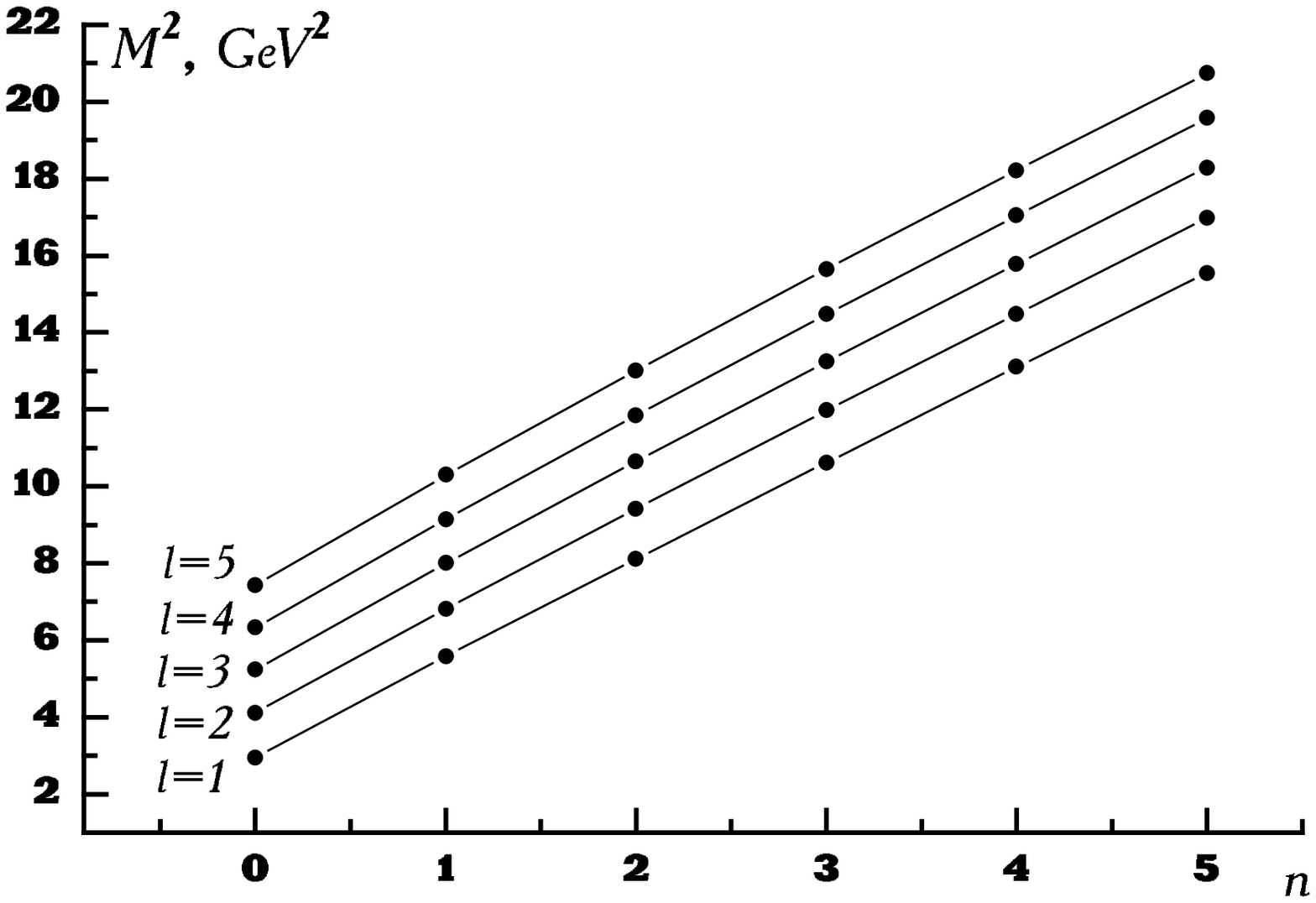,width=10cm}}
\caption{The Regge trajectories for the Hamiltonian (\ref{Hme}) for $m=0$ and
$\sigma=0.17~GeV^2$ (see Table \ref{T5}). Theoretical prediction
for the $\rho$-meson mass ($M^2_{\rho}\approx 1.7~GeV^2$; see the column in Table 
\ref{T4} for $n=0$) is shown not to violate the
straight-line behaviour of the leading theoretical trajectory (see left plot).}
\end{figure}

\begin{figure}[ht]
\centerline{\epsfig{file=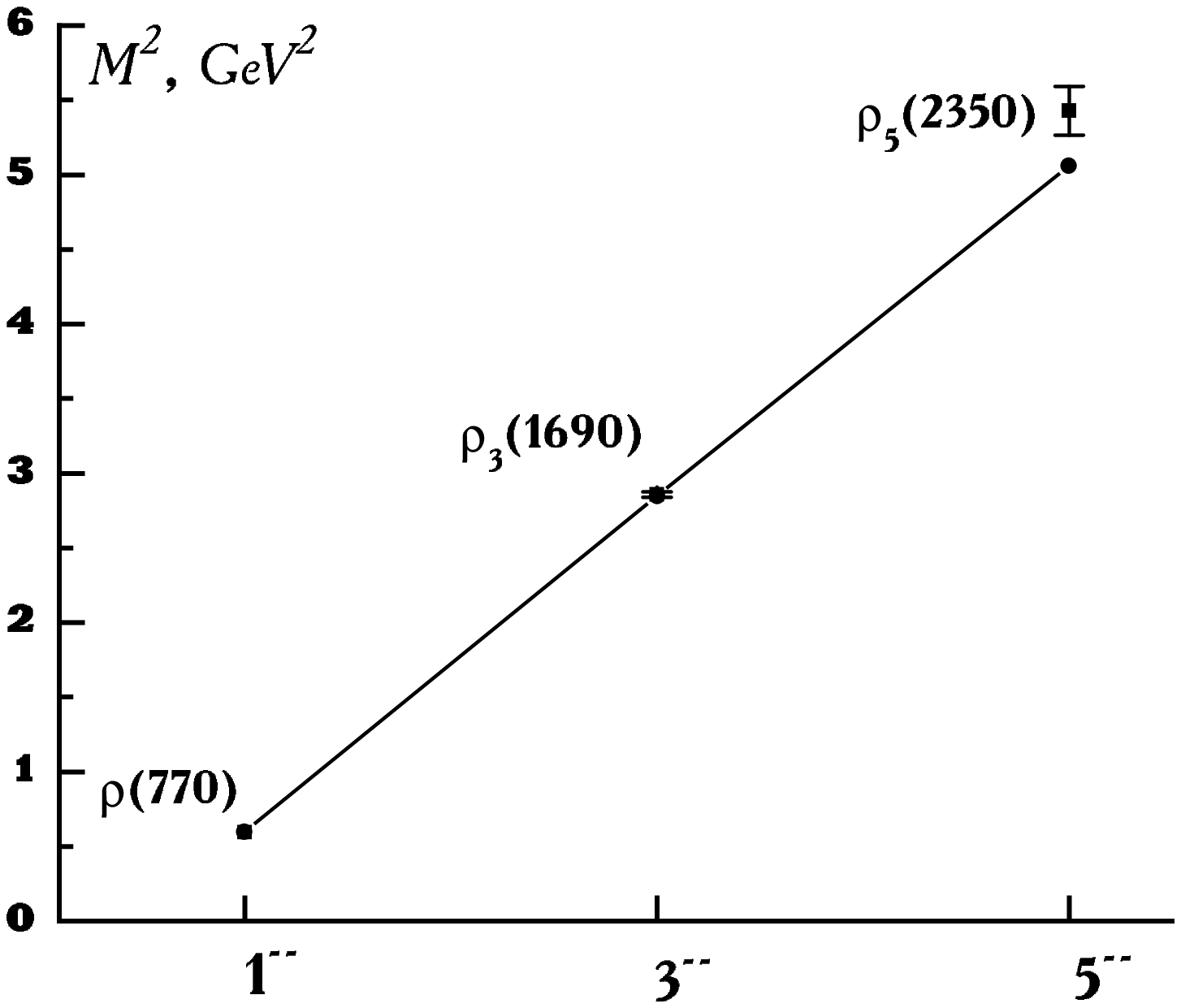,width=10cm}\hspace{-1.5cm}
            \epsfig{file=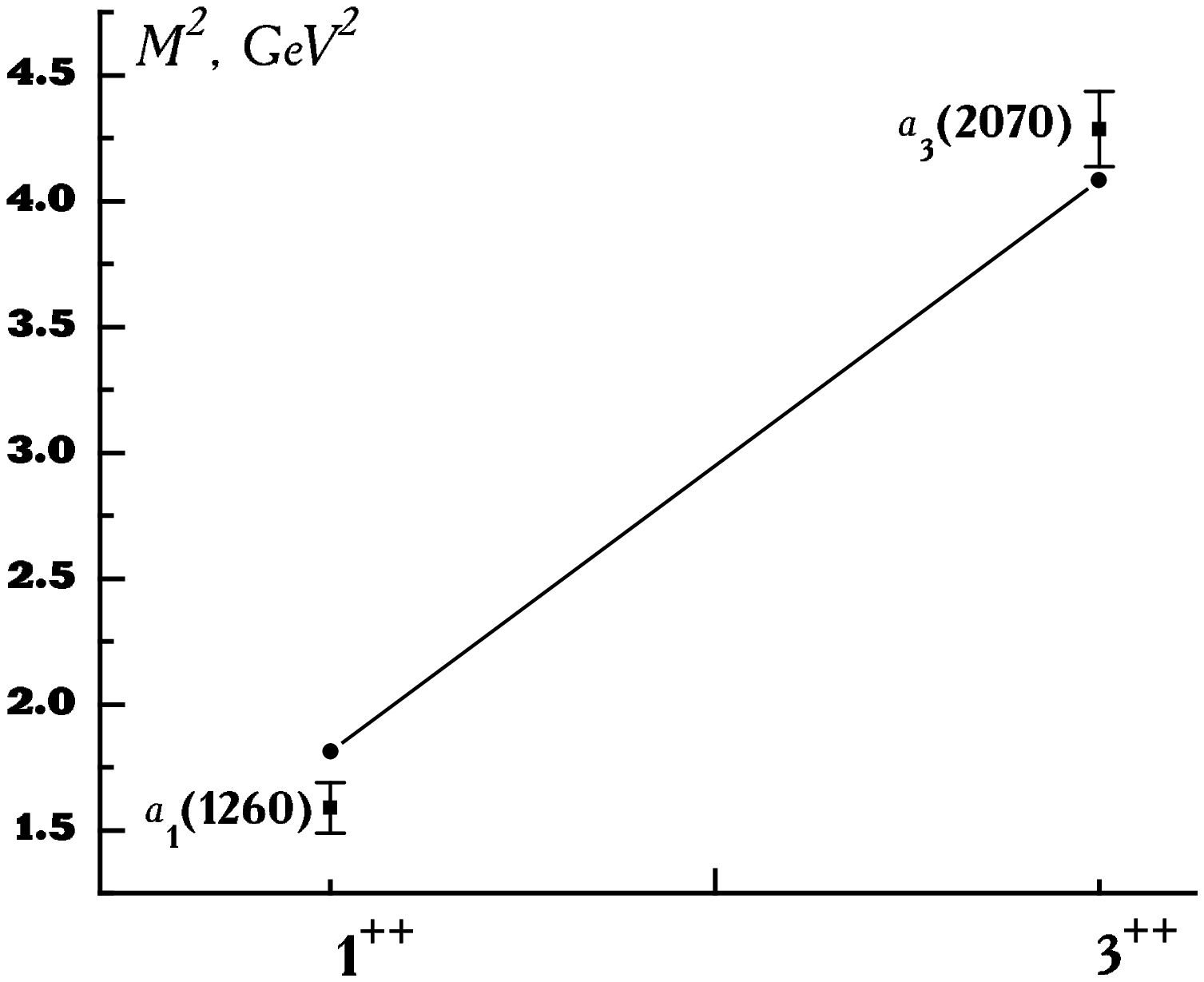,width=10cm}}
\vspace{-1.5cm}
\centerline{\epsfig{file=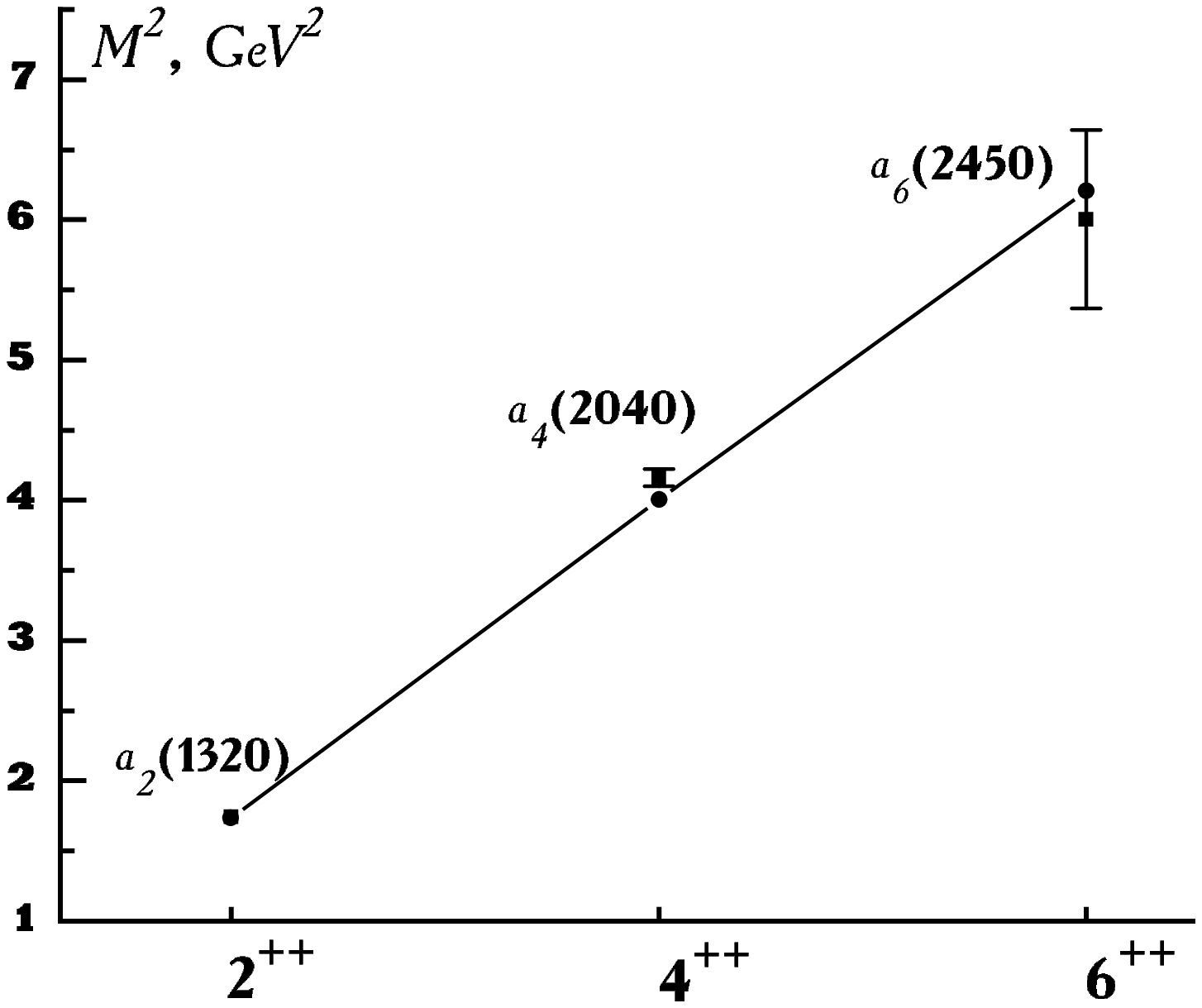,width=10cm}\hspace{8.5cm}}
\caption{The lowest Regge trajectories for light--light mesons fitted with
respect to the 
overall negative mass shift (see Table~\ref{T55}). The theoretical values
for $m=0$ and $\sigma=0.17~GeV^2$ are marked with dots; the experimental data
are given by boxes with error bars.}
\end{figure}
\end{document}